\documentclass[twocolumn]{aastex631}

\usepackage{amsmath}
\usepackage{CJK}
\usepackage{threeparttable} 
\usepackage{booktabs}
\usepackage{threeparttable}
\usepackage{placeins}

\newcommand{\rh}{$r_\mathrm{h}$}

\newcommand{\rev}[1]{#1}
\newcommand{\revmath}[1]{#1}

\providecommand\aj{Astron.\ J.} 
\providecommand\aap{Astron.\ Astrophys.} 
\providecommand\aaps{Astron.\ Astrophys.\ Suppl.} 
\providecommand\apj{Astrophys.\ J.} 
\providecommand\apjl{Astrophys.\ J.} 
\providecommand\apjs{Astrophys.\ J.\ Suppl.} 
\providecommand\icarus{Icarus} 
\providecommand\pasp{PASP} 
\providecommand\mnras{Mon.\ Not.\ R.\ Astron.\ Soc.} 
\providecommand\planss{Planet.~Space~Sci.} 
\providecommand\psj{Planet.\ Sci.\ J.} 
\providecommand\ssr{Space~Sci.~Rev.} 
\providecommand\rnaas{RNAAS}
\providecommand\bain{Bull. Astron. Inst. Netherlands}

\shorttitle{Comet Brightening Model}
\shortauthors{Holt et al.}

\begin{document}
\begin{CJK*}{UTF8}{gbsn}

\title{Empirical Brightening Model for Distant Long-Period Comets}

\correspondingauthor{Carrie E. Holt}

\author[0000-0002-4043-6445]{Carrie E. Holt}
\altaffiliation{LSST-DA Catalyst Postdoctoral Fellow}
\affiliation{Las Cumbres Observatory, 6740 Cortona Drive Suite 102, Goleta, CA 93117, USA}
\affiliation{Department of Astronomy \& the DIRAC Institute, University of Washington, 3910 15th Ave NE, Seattle, WA 98195, USA}
\email{ceholt@uw.edu}

\author[0000-0001-9328-2905]{Colin Snodgrass}
\affiliation{Institute for Astronomy, University of Edinburgh, Royal Observatory, Edinburgh, EH9 3HJ, UK}

\author[0000-0002-3818-7769]{Tim Lister}
\affiliation{Las Cumbres Observatory, 6740 Cortona Drive Suite 102, Goleta, CA 93117, USA}

\author[0000-0002-6702-7676]{Michael S.~P. Kelley}
\affiliation{Department of Astronomy, University of Maryland, College Park, MD 20742, USA}

\author[0000-0003-2781-6897]{Matthew M. Knight}
\affiliation{Volgenau Department of Physics, United States Naval Academy, 572C Holloway Rd, Annapolis, MD 21402, USA}
\affiliation{Department of Astronomy, University of Maryland, College Park, MD 20742, USA}

\author[0000-0002-4838-7676]{Quanzhi Ye (叶泉志)}
\affiliation{Department of Astronomy, University of Maryland, College Park, MD 20742, USA}
\affiliation{Center for Space Physics, Boston University, 725 Commonwealth Ave, Boston, MA 02215, USA}

\author[0000-0003-4507-9384]{Abbie Donaldson}
\affiliation{Institute for Astronomy, University of Edinburgh, Royal Observatory, Edinburgh, EH9 3HJ, UK}

\author[0000-0002-9298-7484]{Cyrielle Opitom}
\affiliation{Institute for Astronomy, University of Edinburgh, Royal Observatory, Edinburgh, EH9 3HJ, UK}

\author[0000-0002-2668-7248]{Dennis Bodewits}
\affiliation{Physics Department, Edmund C. Leach Science Center, Auburn University, Auburn, AL 36849, USA}

\author[0000-0001-7225-9271]{Henry H.\ Hsieh}
\affiliation{Planetary Science Institute, 1700 East Fort Lowell Rd., Suite 106, Tucson, AZ 85719, USA}

\collaboration{42}{(The LCO Outbursting Objects Key (LOOK) Project)}

\author[0000-0001-5749-1507]{Edward Gomez}
\affil{Las Cumbres Observatory, School of Physics and Astronomy, Cardiff University, Queens Buildings, The Parade, Cardiff CF24 3AA, UK}
\affil{Canonical UK Limited, 5 New Street Square, London EC4A 3TW, UK}

\author[0000-0002-1278-5998]{Joey Chatelain}
\affil{Las Cumbres Observatory, 6740 Cortona Drive Suite 102, Goleta, CA 93117, USA}

\author[0000-0002-4439-1539 ]{Sarah Greenstreet}
\affil{Rubin Observatory/NSF's NOIRLab, 950 N. Cherry Ave, Tucson, AZ 85719, USA}
\affil{Department of Astronomy \& the DIRAC Institute, University of Washington, 3910 15th Ave NE, Seattle, WA 98195, USA}



\begin{abstract}
Recent studies have shown the traditional comet brightening model, $Hn$, with a constant activity index $n$ (i.e., single slope in $\log_{10}(r_{\mathrm{h}})$), fails to accurately describe heliocentric lightcurves at large distances, where water-ice sublimation is inefficient. We present a new empirical model, $Hab$, for distant comet brightening in which the effective brightening slope varies with heliocentric distance, $r_{\mathrm{h}}$. Using a multi-year pre-perihelion dataset from uniform instrumentation for 23 comets, we found that $Hab$ provides the best match to long inbound arcs spanning a wide range in $r_{\mathrm{h}}$. The best-fit coefficients $a$ and $b$ cluster tightly for our best-observed comets, suggesting a common brightening behavior; we therefore also tested one- and two-parameter $Hab$ variants in which $a$ and/or $b$ are fixed to anchor values derived from these comets ($A_1 = 0.40 \pm 0.09$, $B_1 = -1.78 \pm 0.68$), either individually, jointly, or via a linear relation. The three-parameter $Hab$ model is strongly favored by the Bayesian information criterion (BIC) for the full lightcurve, justifying its added complexity for characterization. We also tested predictive performance against $Hn$. For brightness forecasting from limited initial data ($\Delta r_{\mathrm{h}} \lesssim 2.5$~au, the heliocentric distance range spanned by observations), $Hab$ itself predicts poorly, whereas a one-parameter reduction, with both coefficients fixed, predicts more reliably, including outperforming $Hn$ with a fixed $n \sim 2$--$4$. Two-parameter reductions with only one coefficient fixed typically reach predictions within 1~mag of the observed brightness after only $\Delta r_{\mathrm{h}} \sim 1.0$~au of inbound coverage. 
\end{abstract}



\section{Introduction} \label{sec:intro}

Long-period comets (LPCs) are some of the most pristine objects in our solar system. Spending most of their lifetime in the Oort cloud until perturbed inward by passing stars or galactic tides, they are thought to have undergone minimal processing, retaining much of their original volatile inventory. Therefore, these objects are valuable probes for connecting observed cometary properties to conditions in the pre-solar nebula and its subsequent evolution.

Thanks to powerful sky surveys, comets today are routinely discovered at heliocentric distances ($r_h$) greater than 5 au, and discoveries of comets at $r_\mathrm{h} > 10$ au are increasing \citep[see Figure 1 in ][]{lister22}. These objects have a visible coma, so we know they are active in some manner. However, the drivers of said activity are not well understood. H$_2$O is the most abundant volatile in most comets and the main driver of cometary activity at $r_\mathrm{h}<3$ au, but water-ice sublimation is inefficient at $r_\mathrm{h}>5$ au \citep{meech04}. Distant activity is likely driven by more volatile ices like CO and CO$_2$, which are next in abundance and can lift grains at heliocentric distances of 6--8 au for CO$_2$ and tens of au for CO \citep{meech04,jewitt21}.  Indeed, CO and CO$_2$ have been detected in comets and Centaurs beyond 5 au \citep[e.g.,][]{ yang21, harringtonpinto23, snodgrass25, roth25, schambeau26}. The one-time transition from amorphous to crystalline water-ice has also been suggested to drive activity from $r_\mathrm{h}\sim3$~au up to $r_\mathrm{h}\sim10$~au \citep[e.g.,][]{smoluchowski81,jenniskens96, prialnik22}, and has been argued as an explanation for Centaur activity \citep[cf.][]{jewitt09a,guilbertlepoutre12}.

While spectroscopy and narrowband imaging provide more direct constraints on activity drivers, they generally require higher signal-to-noise per exposure and more specialized setups (e.g., CO$_2$ can only be directly measured using space-based observatories), making them less practical for sustained, high-cadence monitoring of large samples of faint distant comets. Broadband optical photometry is the most accessible means of investigating distant cometary activity for large samples of targets, allowing us to monitor brightness changes as comets approach the Sun and relate those changes to potential activity drivers. Broadband photometric monitoring observations can be carried out on small ($\sim$1-2m), ground-based telescopes, which can allocate substantially more observing time across many nights than is feasible with larger-aperture telescopes. Detailed physical models can be used to connect the observed broadband brightness of a comet to the rate of outgassing of the driving species \citep[e.g.][]{meech04}, but these require knowledge of (or more often assumptions on) a wide range of parameters, including nucleus size, dust-to-gas ratios, dust size distribution and reflectance properties. More commonly, the evolution in brightness of comets is described using empirical relationships, that allow comparison of relative brightness corrected for observing geometry, but are agnostic to the detailed physics of the activity process. The slope of the brightening rate has been used to characterize LPCs since the early development of the Oort-cloud concept \citep{oort51,whipple78,ahearn95}. 

\subsection{The Traditional Model (Hn)}
Characterizing the brightness evolution of comets captured by broadband photometry requires a functional form relating apparent brightness to heliocentric and geocentric distance. A commonly used reference normalization to ensure the most accurate comparisons for photometric magnitudes is $H(1,1,0)$, representing the theoretical brightness for a Sun-based observer with the comet at $r_\mathrm{h} = 1$~au (thus with a comet-observer distance of 1~au and a phase angle of $\alpha=0^{\circ}$). All $H(1,1,0)$ values in this work are computed in $r$-band. Hereafter we refer to this $H(1,1,0)$ as $H$. Traditionally for comets, a parameter $n$, also known as the activity index, has been used to characterize the brightening rate of LPCs \citep{everhart67a}, so that the apparent magnitude $m_\mathrm{app}$ can be described as: 

\begin{equation}
\label{eqn:traditional}
\begin{split}
m_\mathrm{app} = {}& H(1,1,0) + 5\log_{10}(\Delta) \\
& + 2.5n\log_{10}(r_\mathrm{h}) - 2.5\log_{10}(\phi(\alpha))
\end{split}
\end{equation}

\noindent We refer to Equation~\ref{eqn:traditional} as the \emph{traditional model} ($Hn$) with two free parameters ($H$ and $n$), where $\Delta$ is the geocentric distance in au; 
the coefficient 5 reflects the $1/\Delta^2$ dependence of flux on distance ($2 \times 2.5 = 5$). 
$r_h$ is the heliocentric distance in au, $\alpha$ is the phase angle (the Sun-Target-Observer angle), and $\phi(\alpha)$ is the phase function, which is the ratio of the brightness at phase angle $\alpha$ to that at $\alpha=0^{\circ}$. To calculate $\phi(\alpha)$, we used the Schleicher-Marcus phase function for cometary dust\footnote{\url{http://asteroid.lowell.edu/comet/dustphase.html}},  which is a composite of the phase function from comet Halley \citep{schleicher98} for smaller phase angles and the phase function derived for near-Sun comets in \citet{marcus07b,marcus07c} at mid- to large-phase angles \citep{schleicher11}. 

Today, Equation \ref{eqn:traditional} is the most commonly used model for brightening rate characterization and predicting future brightness. Commonly used resources for comet magnitude estimates, including JPL Horizons\footnote{\url{https://ssd.jpl.nasa.gov/horizons/}} and the Minor Planet Center, adopt this model, fitting the data using least squares. In JPL brightness calculations, the absolute magnitude ($M_1$) and magnitude slope parameter ($K_1$) correspond to Equation~\ref{eqn:traditional} as $H(1,1,0) = M_1$ and $n = K_1/2.5$. These values are for the total magnitude, including the coma. For some comets, JPL brightness calculations may also include $M_2$ and $K_2$ parameters, which correspond to the magnitude and the slope of the central condensation (or pseudo-nucleus) alone.

\subsection{A Fixed-Aperture Scaling Correction}
\label{sec:intro_aperture}
A fixed angular photometric aperture corresponds to a smaller physical radius at the comet as it approaches the Earth (smaller $\Delta$), so a growing fraction of the coma falls outside the aperture. This partially offsets the inverse-square brightening a point source would show, so a $\Delta^{-1}$ scaling is more appropriate than the point-source $\Delta^{-2}$ scaling in Equation~\ref{eqn:traditional} \citep{opik63, jewitt91}. Because our photometry uses a fixed $5\arcsec$ aperture, we adopted $2.5\log_{10}(\Delta)$ in place of the $5\log_{10}(\Delta)$ term for all model fits in this work:
\begin{equation}
\label{eqn:aper_correct}
\begin{split}
m_\mathrm{app} = {}& H(1,1,0) + 2.5\log_{10}(\Delta) \\
& + 2.5n\log_{10}(r_\mathrm{h}) - 2.5\log_{10}(\phi(\alpha))
\end{split}
\end{equation}
We discuss the limitations of this choice in Section~\ref{sec:limitations}.

\subsection{Limitations of the Traditional Model}
\label{sec:intro_limitations}

Within 3~au, the $Hn$ model has also been used to compare brightening behavior across dynamical classes of LPCs, particularly between dynamically new comets (DNCs) and returning comets. DNCs are a subset of LPCs distinguished by their original orbital energy, proportional to $1/a_0$, the inverse of the original semi-major axis calculated with respect to the solar system barycenter, found by integrating the orbit backward until planetary perturbations become negligible ($r_\mathrm{h}\sim250$~au). The distribution of $1/a_0$ shows an excess of comets with $a_0 > 10^4$~au, a ``spike'' that motivated the postulation of the Oort cloud \citep{oort50, oort51}. However, there is no definitive threshold value of $1/a_0$ separating new from returning comets, and recent work \citep{dybczynski15, dybczynski22a} has shown that not all comets in the Oort spike are truly dynamically new (and therefore compositionally pristine). In practice, a common threshold of $1/a_0 < 4\times10^{-5}$~au$^{-1}$ has been proposed to identify DNCs \citep{krolikowska17}, though the exact value varies between studies.

Historically, DNCs have been observed to have shallower brightening slopes than returning comets \citep{whipple78}, a result recently confirmed by \citet{lacerda25} using a large collection of Minor Planet Center photometry. To account for heterogeneity between observers and instruments, \citet{lacerda25} grouped observations into ``strands,'' consecutive measurements from a single observatory and filter, and applied quality cuts before fitting each strand independently, isolating relative brightening trends from absolute calibration differences between sources. \citet{lacerda25} found median values of $K_1$ of 6.7, 7.8, and 13.4 ($n = 2.7, 3.1, 5.4$) for new, intermediate, and old LPCs (i.e., $1/a_0$ values of 1, 20, and 400~$\times10^{-4}$), respectively, on their inbound legs where $r_\mathrm{h}<3$~au. With modern telescopes and sky surveys now routinely discovering comets beyond $r_\mathrm{h}=3$~au, where water-ice sublimation becomes inefficient and more volatile species are thought to drive activity, it has become possible to test whether the traditional model remains an adequate description at these distances. Previous broadband photometric studies provided early insight into this regime, but either lacked comprehensive heliocentric distance coverage \citep{sarneczky16,meech09a} or focused on individual objects \citep{womack17}. 

\begin{figure*}[ht]
    \centering
    \includegraphics[width=0.75\textwidth]{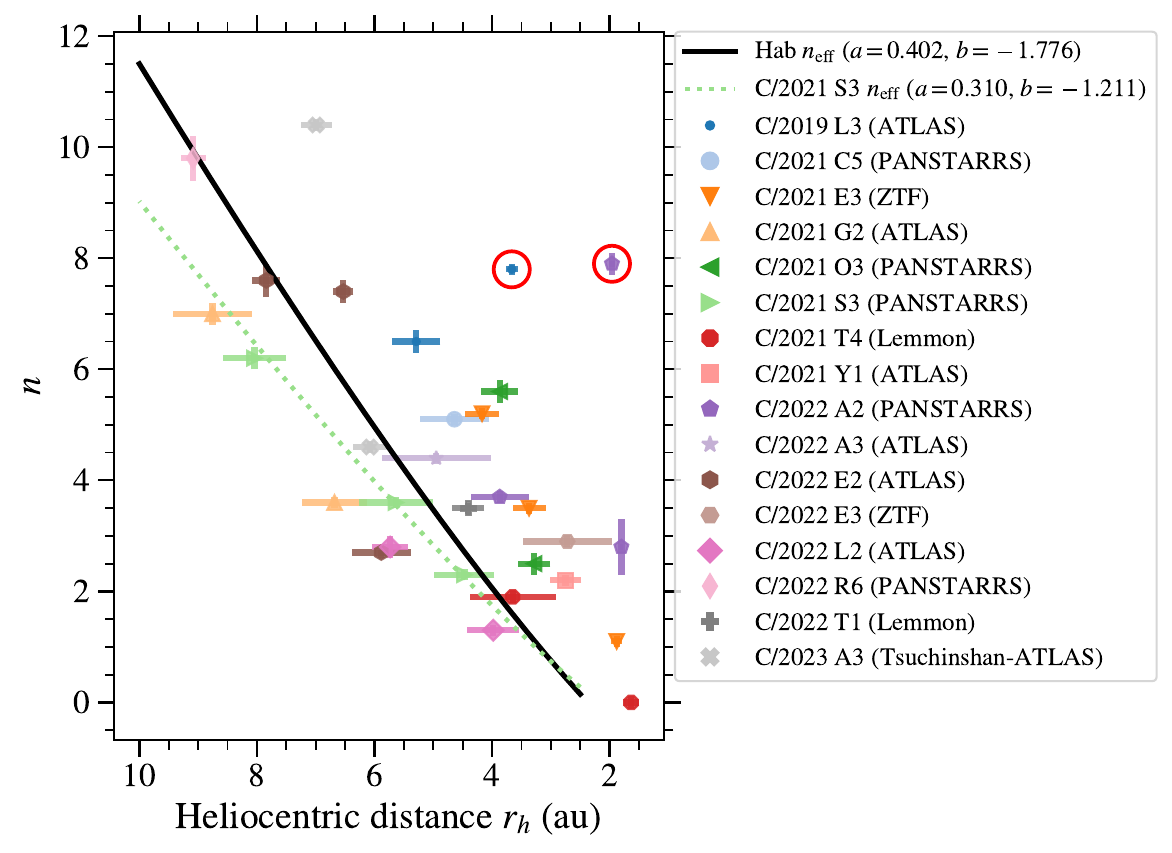}
    \caption{Activity index $n$ for long-period comets, where each color and marker corresponds to a different comet and has a changing $n$ value for piece-wise linear fits over different heliocentric ranges across a comet's lightcurve, adapted from Fig.~6 of \citet{holt24}. The horizontal error bar for each point corresponds to the heliocentric distance range over which the $n$ was fit and the vertical error bar corresponds to the uncertainty in the fit, and errorbars not visible are smaller than the plotted points. The black line, discussed further in Section \ref{sec:3param}, is the effective slope of the three-parameter model ($n_\mathrm{eff}$; Equation~\ref{eqn:effective_n}) using the fixed reference values $A_1 = 0.40$ and $B_1 = -1.78$ as $a$ and $b$. The light green dotted line shows the $n_\mathrm{eff}$ of the three-parameter model fit to comet C/2021 S3, whose multiple $n$ values are the light green points. Two lightcurve segments that fall outside of this trend, near $n \approx 8$ within $r_\mathrm{h} < 4$~au (circled in red), belong to C/2019 L3 and C/2022 A2, which experienced a ``surge'' in brightness near perihelion.}
    \label{fig:holt24fig}
\end{figure*}

Recent multi-object, multi-epoch studies have begun to characterize the brightening behavior of LPCs beyond $r_\mathrm{h}=3$~au. \citet{holt24} used uniformly processed broadband photometry of 21 LPCs, finding that $n$ systematically decreases with decreasing \rh\ for individual comets and across the population as a whole (Figure \ref{fig:holt24fig}; discussed further in Section \ref{sec:3param}), with no clear difference between DNCs and returning comets, though the sample included only three returning comets. \citet{lacerda25} further showed that the brightening slope of new and intermediate comets varies significantly at $r_\mathrm{h}>3$~au, while old comets show no such trend. It is worth noting that the two studies used different $1/a_0$ thresholds to define DNCs: \citet{holt24} adopted the threshold of $1/a_0 < 4 \times 10^{-5}$~au$^{-1}$ from \citet{krolikowska17} (which we also adopt here), while \citet{lacerda25} followed the Oort \& Schmidt grouping with $1/a_0 < 10^{-4}$~au$^{-1}$ for new comets \citep{oort51}. Across both studies, a constant $n$ proves insufficient to describe the full inbound lightcurve, and post-perihelion behavior at large distances ($r_\mathrm{h} \gtrsim 5$~au) remains poorly characterized.

\subsection{This Work}

The inadequacy of a constant $n$ at large heliocentric distances presents practical problems when using the $Hn$ model to predict the distant behavior of comets, such as attempting to assess the likely yield of new surveys (e.g., LSST, Donaldson et al., in prep; NEOSurveyor, \citealt{levine25}), or to extrapolate the observed brightness of a comet found at large distances to predict its perihelion activity level.  The latter is important for the European Space Agency mission Comet Interceptor \citep{jones24}, which is scheduled to launch between August 2028 and July 2029, and will target a yet-to-be-discovered LPC. Based on initial observations of a comet discovered far from the Sun (ideally at $r_h > 10$ au, to give sufficient warning time to have the spacecraft position itself for a flyby) the mission team must make predictions on the expected evolution of the comet's activity level to select a good target: a comet with too high levels of dust production may be unsafe for the spacecraft, while a weakly outgassing comet may not provide enough material for in situ instruments to sample properly \citep{dekeyser2025,snodgrass-roma}.

A $r_\mathrm{h}$-dependent model is needed to improve accuracy in both characterization and predictions. In this paper, we extended the analysis of \citet{holt24} using recent observations of comets discovered at $r_\mathrm{h} > 5$~au to develop the $Hab$ model, an empirical brightening model in which the activity index varies with heliocentric distance. We test how well the $Hab$ model fits the full data sets relative to the traditional model ($Hn$), and whether the tight clustering of its best-fit parameters justifies simpler, constrained variants with fewer free parameters. We then compare the predictive performance of all of these models from limited early data. Section \ref{sec:data} describes the data used to fit the model. Section \ref{sec:brightening_models} describes the $Hab$ model and its variants, and Section \ref{sec:methods} describes the testing strategies. The results are described in Section \ref{sec:results}, followed by a discussion of the implications and caveats in Section \ref{sec:discussion}.

\section{Data}
\label{sec:data}
The dataset used in this study is an extension of the one presented in \citet{holt24}. Specifically, we added eight additional comets to the 15 LPCs from the original study with observational coverage spanning a range of heliocentric \rev{distance} of at least 1~au ($\Delta r_\mathrm{h} > 1$~au), while still pre-perihelion and beyond $r_\mathrm{h}=2.5$~au, for a total of 23 comets (Table \ref{tab:look_sample}). We also extended the pre-perihelion data arcs for seven of the original 15 comets. Observational parameters and data reduction procedures remain unchanged from \citet{holt24}, where they are described in full.

As in our previous work, we leverage Las Cumbres Observatory's (LCO) global network \citep{brown13} of 1~m robotic telescopes as part of the LCO Outbursting Objects Key (LOOK) Project \citep{lister22}. This network provides uniform instrumentation across both hemispheres and enables frequent monitoring (here typically every three days) over extended periods (usually years). Observations were taken using the Sloan Digital Sky Survey (SDSS) $g^\prime$ and $r^\prime$ filters, photometrically calibrated to the Pan-STARRS1 system using \texttt{calviacat} \citep{kelley19} and the ATLAS-RefCat2 catalog \citep{tonry12}, with color corrections derived from LOOK Project photometry; photometry calibrated to the PS1 system using AB magnitudes is hereafter denoted $g$ and $r$. Since 2020, we have observed all comets discovered inbound at $r_\mathrm{h} \gtrsim 5$~au with a heliocentric distance range of $\Delta r_\mathrm{h} > 1$~au before perihelion. Our strategy allows for detailed characterization of brightness evolution as comets approach the Sun from beyond $r_\mathrm{h}=5$~au. Because the LCO network enables observations from both the northern and southern hemispheres, the only major observational gaps occur during solar conjunction.

For our data reduction, we applied the same image quality cuts described in \citet{holt24}: excluding images with point-source full widths at half maximum (FWHM) greater than 4.0$''$ or photometric uncertainties exceeding 0.15 mag. Images were also manually inspected and removed from consideration if affected by background source contamination. Photometric calibration, including standard color corrections, followed the procedures established in that study as well. Finally, we adopted the same justification for using a 5\arcsec{} photometric aperture, which provides the most consistent results for photometric monitoring across a broad range of heliocentric distances.

In this study, we focused on the pre-perihelion evolution and predictive modeling of cometary brightness. To isolate activity unrelated to water-ice sublimation, our analysis used only observations from $r_\mathrm{h} > 2.5$~au. This cutoff was motivated by the findings of \citet{holt24}, who showed that nearly all comets in the sample with perihelion distance $q < 2.5$~au exhibited a sharp increase in brightening at $r_\mathrm{h}<2.5$~au, likely due to the onset of water-driven activity. All comets in our sample were observed pre-perihelion through at least our 2025 May 1 cutoff; some objects had additional pre-perihelion data available beyond this date that are not included here. We reserve analysis of the $q < 2.5$~au region, including post-perihelion data, for future work.

For our statistical tests, we first focused on the seven comets with the largest ratio of maximum to minimum heliocentric distance ($\mathcal{R} \equiv r_\mathrm{h,max} / r_\mathrm{h,min}$), since a larger $\mathcal{R}$ provides greater leverage for discriminating between models in logarithmic space. Because our analysis is restricted to observations at $r_\mathrm{h} > 2.5$~au, and a comet can never be observed pre-perihelion below its own perihelion distance $q$, $r_\mathrm{h,min}$ is bounded below by 2.5~au, or by $q$ itself for comets with $q > 2.5$~au. The maximum possible $\mathcal{R}$ for a given comet is therefore $r_\mathrm{h,max}$ divided by whichever of these two limits is larger. A comet's actual $\mathcal{R}$ falls short of this ceiling whenever observations stopped before reaching this limit, for example due to a gap during solar conjunction. These seven high-$\mathcal{R}$ comets have data starting at $r_\mathrm{h} = 6$--$9$~au with $\mathcal{R} > 2$. We then expanded our analysis to include the full sample of 23 comets. For testing the predictive power of the models, we used those same seven comets that have $\mathcal{R} > 2$. Table~\ref{tab:look_sample} details the comets used for our analyses.

\urldef{\nakanourl}\url{https://www.oaa.gr.jp/~oaacs/nk.htm}
\begin{table*}
\centering
\caption{Comet Sample Sorted by $\mathcal{R}$}
\label{tab:look_sample}
\footnotesize
\setlength{\tabcolsep}{0.08in}
\renewcommand{\arraystretch}{0.9}
\begin{tabular}{lcccclccc}
\hline\hline
Comet Name & $r_{\mathrm{h,max}}$ & $\mathcal{R}$ & $\Delta r_{\mathrm{h}}$ & $q$ & T & $N$ & $g-r$ & $1/a_0 \times 10^{6}$ \\
 & (au) & & (au) & (au) & & & (mag) & (au$^{-1}$) \\
\hline
C/2023 A3 (Tsuchinshan-ATLAS) & 7.25 & 2.83 & 4.69 & 0.39 & 2024 Sep 27 & 162 & 0.51 & 7 \\
C/2020 R7 (ATLAS) & 6.93 & 2.34 & 3.97 & 2.95 & 2022 Sep 15 & 185 & 0.51 & 10 \\
C/2021 T4 (Lemmon) & 6.73 & 2.30 & 3.81 & 1.48 & 2023 Jul 31 & 58 & 0.50 & 34 \\
C/2022 L2 (ATLAS) & 6.04 & 2.24 & 3.35 & 2.69 & 2024 Mar 11 & 117 & 0.51 & -1 \\
C/2022 E2 (ATLAS) & 8.08 & 2.20 & 4.42 & 3.67 & 2024 Sep 14 & 179 & 0.48 & 72 \\
C/2021 S3 (PANSTARRS) & 8.58 & 2.16 & 4.62 & 1.32 & 2024 Feb 14 & 140 & 0.56 & -1 \\
C/2023 C2 (ATLAS) & 5.97 & 2.14 & 3.18 & 2.37 & 2024 Nov 16 & 38 & 0.51 & 54 \\
\hline
C/2022 S4 (Lemmon) & 6.44 & 1.99 & 3.21 & 2.76 & 2024 Jul 18 & 126 & 0.52 & 73 \\
C/2021 G2 (ATLAS) & 9.42 & 1.89 & 4.43 & 4.98 & 2024 Sep 9 & 208 & 0.49 & 23 \\
C/2024 E1 (Wierzchos) & 7.71 & 1.82 & 3.47 & 0.57 & 2026 Jan 20 & 64 & 0.50 & 28 \\
C/2019 L3 (ATLAS) & 5.70 & 1.60 & 2.15 & 3.56 & 2022 Jan 10 & 77 & 0.48 & 26 \\
C/2022 QE78 (ATLAS) & 8.73 & 1.56 & 3.14 & 5.48 & 2025 Sep 12 & 57 & 0.49 & 15 \\
C/2022 A3 (ATLAS) & 5.88 & 1.54 & 2.07 & 3.71 & 2023 Sep 29 & 98 & 0.51 & 1148 \\
C/2023 R1 (PANSTARRS) & 8.11 & 1.47 & 2.61 & 3.57 & 2026 Apr 13 & 88 & 0.50 & 6 \\
C/2024 A1 (ATLAS) & 5.75 & 1.47 & 1.84 & 3.87 & 2025 Jun 12 & 87 & 0.52 & 332 \\
C/2022 E3 (ZTF) & 3.75 & 1.47 & 1.19 & 1.12 & 2023 Jan 12 & 24 & 0.54 & 713 \\
C/2023 T3 (Fuls) & 5.18 & 1.46 & 1.63 & 3.55 & 2025 Jan 25 & 53 & 0.50 & 1648 \\
C/2021 E3 (ZTF) & 4.46 & 1.45 & 1.37 & 1.78 & 2022 Jun 10 & 37 & 0.48 & -11 \\
C/2022 R6 (PANSTARRS) & 9.31 & 1.40 & 2.65 & 6.57 & 2025 Aug 27 & 177 & 0.50 & 21 \\
C/2021 O3 (PANSTARRS) & 4.18 & 1.39 & 1.17 & 0.29 & 2022 Apr 21 & 24 & 0.48 & -42 \\
C/2022 U3 (Bok) & 6.70 & 1.36 & 1.78 & 4.83 & 2024 Jul 28 & 53 & 0.56 & 25 \\
C/2022 T1 (Lemmon) & 4.68 & 1.36 & 1.24 & 3.44 & 2024 Feb 17 & 43 & 0.52 & 11 \\
C/2021 C5 (PANSTARRS) & 5.23 & 1.29 & 1.18 & 3.24 & 2023 Feb 11 & 46 & 0.47 & 16 \\
\hline
\multicolumn{9}{p{0.9\textwidth}}{$r_\mathrm{h,max}$: heliocentric distance of the first observation with LOOK. $\mathcal{R} \equiv r_\mathrm{h,max}/r_\mathrm{h,min}$: ratio of the largest to smallest heliocentric distance. $\Delta r_\mathrm{h}$: pre-perihelion heliocentric distance range with $r_\mathrm{h} > 2.5$~au. $q$: perihelion distance. T: perihelion date. $N$: number of binned observations as of 2025 May 1; observations were binned by night and by filter, so the number of visits is approximately $N/2$. $g-r$: median $g-r$ color using a 5$''$ aperture. $1/a_0$: reciprocal original semi-major axis from the Nakano Notes (\nakanourl), used as a proxy for dynamical age. The horizontal line separates the seven comets with $\mathcal{R} > 2$ (above) from the rest of the sample.}\\
\hline
\end{tabular}
\end{table*}

\section{An Empirical Brightening Model}
\label{sec:brightening_models}

\subsection{The \texorpdfstring{$Hab$}{Hab} Model}
\label{sec:3param}

\begin{figure}[tbp]
    \centering
    \includegraphics[width=\columnwidth]{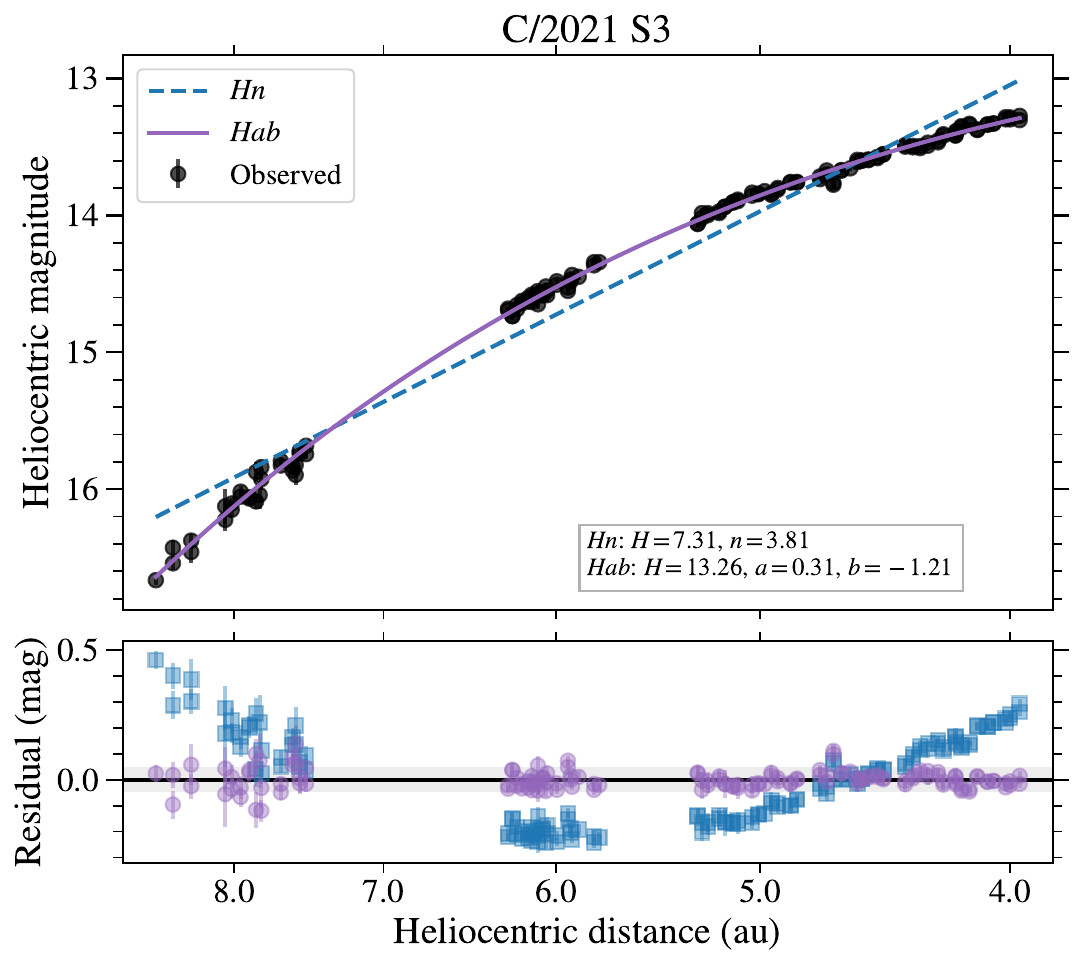}
    \caption{Pre-perihelion heliocentric lightcurve of C/2021 S3 in $r$-band. The upper panel shows the variation of pre-perihelion heliocentric magnitude with heliocentric distance using a 5\arcsec{} aperture. The black points represent the $r$ and $g$ magnitudes, where the $g$ magnitudes are converted to the $r$-band by shifting them by the median $g - r$ color of 0.56; observations in other bands, if available, are not included here. \rev{The lines represent the best fits} for both the $Hn$ model (dashed blue) and $Hab$ model (solid purple), determined by the least-squares method. The lower panel shows the residuals for both model fits. The $x$-axis is in log scale. }
    \label{fig:21s3}
\end{figure}

Motivated by the findings of \citet{holt24} and \citet{lacerda25} that a constant $n$ fails to describe distant lightcurves, we replaced the brightening slope ($n$) in Equation~\ref{eqn:aper_correct} with a linear function of heliocentric distance ($ar_\mathrm{h} + b$), where $a$ and $b$ are independent model parameters. The improvement is visible directly in the residuals: for comet C/2021 S3, one of our best-sampled objects, the $Hab$ fit tracks the curvature of the observed lightcurve, while the single-slope $Hn$ fit does not (Figure \ref{fig:21s3}).

The term $a r_\mathrm{h} + b$ in this model should be interpreted as a heliocentric-distance-dependent activity index rather than as the instantaneous analogue of the constant-$n$ slope in the traditional model. We adopted this form because it provides a simple way to capture the observed curvature of distant lightcurves over a wide range of $r_\mathrm{h}$, while remaining comparable to the traditional constant-$n$ parameterization. In the traditional model, where $n$ is constant, the slope of the lightcurve with respect to $\log_{10} r_\mathrm{h}$ is simply $2.5n$. In contrast, when the activity index varies with $r_\mathrm{h}$, differentiating the magnitude with respect to $\log_{10} r_\mathrm{h}$ introduces additional terms from the $r_\mathrm{h}$-dependence. For comparison with the traditional model, we therefore define an effective slope $n_{\rm eff}$ for the $Hab$ model:
\begin{equation}
\label{eqn:effective_n}
\begin{split}
n_\mathrm{eff} &= a r_\mathrm{h} + b + a r_\mathrm{h} \ln{(10)} \log_{10}{(r_\mathrm{h})} \\
&= a r_\mathrm{h} (\ln{(r_\mathrm{h})} + 1) + b
\end{split}
\end{equation}

\begin{figure*}
    \centering
    \includegraphics[width=0.7\textwidth]{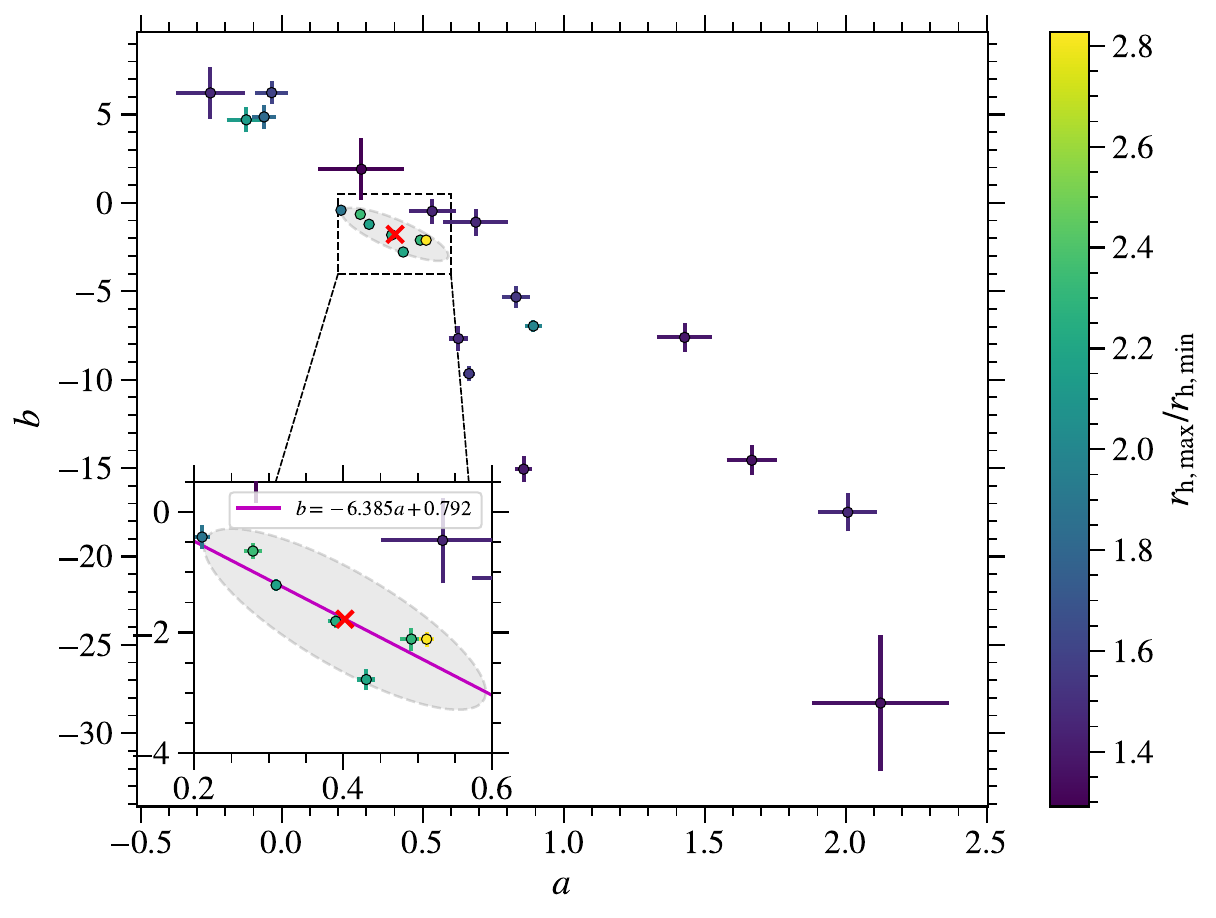}
    \caption{The distribution of $a$ and $b$ parameters with 1$\sigma$ error bars for the $Hab$ model. Points are colored according to the $\mathcal{R} \equiv r_\mathrm{h,max}/r_\mathrm{h,min}$ ratio of the data used to fit the model. In the inset box, we highlight the 6 comets with the largest $\mathcal{R}$. The center of the distribution of the small group is marked by a red $\times$ symbol, and the gray-shaded ellipse shows the 2$\sigma$ standard deviation. A diagonal magenta line shows the linear regression of the small group.}
    \label{fig:threeparamdist}
\end{figure*}

The $Hab$ model has three free parameters, compared to two for the traditional model ($Hn$), which improves flexibility but can also lead to poorly constrained fits when the heliocentric distance ratio sampled, $\mathcal{R}$, is limited. The best-fit $a$ and $b$ parameters cluster in a relatively narrow region for the best-observed comets (Figure~\ref{fig:threeparamdist}): the six comets with the largest $\mathcal{R}$ are well described by a distribution centered on $a = 0.40$ and $b = -1.78$, with $1\sigma$ standard deviations of 0.09 and 0.68, respectively. Hereafter we use $A_1$ and $B_1$ to denote these fixed reference values, distinct from the free parameters $a$ and $b$. These six comets are a subset of the seven $\mathcal{R} > 2$ comets used as the high-quality sample in this work (Table~\ref{tab:look_sample}); the seventh, C/2023~C2, has the smallest $\mathcal{R}$ (2.14) of the group and only 38 observations, and is retained in the broader $\mathcal{R} > 2$ sample used for the predictive tests in Section~\ref{sec:results_predict}. These six comets also appear to have their own linear correlation, which we fit with $b = -6.385a + 0.792$. This clustering suggests that simplified variants with fewer free parameters may capture the general brightening behavior of most LPCs; in Section~\ref{sec:hab_variants} we use $A_1$, $B_1$, and this correlation to construct one- and two-parameter variants of the $Hab$ model. \rev{Part of this correlation may reflect the natural covariance between $a$ and $b$ when fitting over a limited $r_\mathrm{h}$ range rather than a purely physical relationship; we use it as an empirical constraint, and leave a more detailed covariance or pivot-point analysis with a larger sample to future work.} \rev{These anchor values are robust to the choice of comets: recomputing them six times, each time excluding one of the six comets, gives $A_1 = 0.38$ to 0.43 and $B_1 = -2.00$ to $-1.58$, well within their $1\sigma$ spread, and $a$--$b$ relations that remain within the $2\sigma$ population ellipse over the observed range of $a$.}

\subsection{\texorpdfstring{$Hab$}{Hab} Model Variants}
\label{sec:hab_variants}

We constructed four constrained versions of the $Hab$ model (Table \ref{tab:models}): a one-parameter model with both $a$ and $b$ fixed to $A_1$ and $B_1$ (Fixed-$ab$), two two-parameter models with only $a$ fixed to $A_1$ or only $b$ fixed to $B_1$ (Fixed-$a$ and Fixed-$b$, respectively), and a two-parameter model in which $b$ is fixed as a linear function of $a$ (Correlated-$ab$).
\begin{table*}
\centering
\caption{Tested Models}
\label{tab:models}
\footnotesize
\begin{tabular}{llll}
\hline\hline
Name & Activity Term\textsuperscript{a} & Free Parameters & Description\\
\hline
$Hn$ & $n$ & $H$, $n$ & Traditional model; no $r_\mathrm{h}$ dependence\\
$Hab$ & $a r_\mathrm{h} + b$ & $H$, $a$, $b$ & Empirical 3-parameter model\\
\hline
Fixed-$ab$ & $A_1 r_\mathrm{h} + B_1$ & $H$ & Fully constrained: $a$ and $b$ are defined\\
Fixed-$a$ & $A_1 r_\mathrm{h} + b$ & $H$, $b$ & Partially constrained: Only $a$ is defined\\
Fixed-$b$ & $a r_\mathrm{h} + B_1$ & $H$, $a$ & Partially constrained: Only $b$ is defined\\
Correlated-$ab$ & $a r_\mathrm{h} - 6.385a + 0.792$ & $H$, $a$ & Partially constrained: $a$ is correlated to $b$\\
\hline
\multicolumn{4}{p{0.85\textwidth}}{\textsuperscript{a}Replaces $n$ in Equation~\ref{eqn:aper_correct}, but is not equivalent to $n$ (see $n_{\rm eff}$, Equation~\ref{eqn:effective_n}).}\\
\multicolumn{4}{p{0.85\textwidth}}{The four models below the horizontal line are fully or partially constrained variants of the unrestricted 3-parameter $Hab$ model, where $A_1 = 0.40$ and $B_1 = -1.78$.}\\
\hline
\end{tabular}
\end{table*}

The Fixed-$ab$ model, in which both the slope ($a$) and intercept ($b$) of the heliocentric brightening term are fixed to $A_1$ and $B_1$, has only a single free parameter, $H(1,1,0)$. This is analogous to the traditional $Hn$ model with a fixed $n$ (typically assumed to be $n \sim 2$--4), but with a distance-dependent brightening term in place of a constant slope.

The Fixed-$a$ and Fixed-$b$ models instead fix only $a$ to $A_1$, or only $b$ to $B_1$, respectively, leaving the other coefficient free along with $H(1,1,0)$. In the Correlated-$ab$ model, we instead use the correlation found in the distribution of $a,b$ parameters to constrain $b$ as a function of $a$, such that the $Hab$ activity term $ar_\mathrm{h} + b$ becomes $ar_\mathrm{h} - 6.385a + 0.792$. These two-parameter models allow us to assess whether partial flexibility in the brightening behavior improves performance relative to the fully fixed Fixed-$ab$ model, while still avoiding the potential overfitting risk of the full three-parameter $Hab$ model.

\begin{figure*}[t]
\centering
\includegraphics[width=0.6\textwidth]{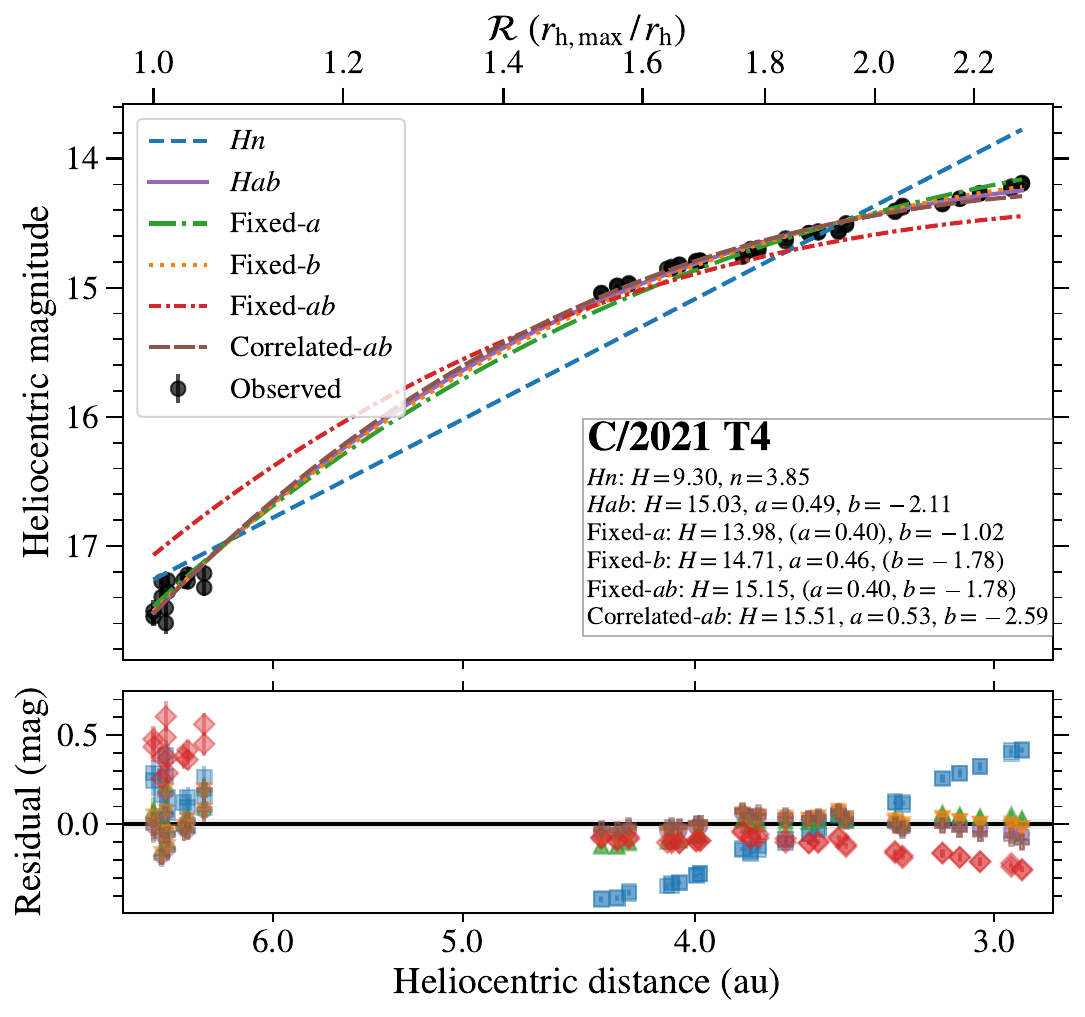}
\caption{Pre-perihelion heliocentric lightcurve of C/2021 T4 in $r$-band. The upper panel shows the variation of pre-perihelion heliocentric magnitude with heliocentric distance using a $5''$ aperture; the $x$-axis is in log scale, with a secondary axis along the top giving the corresponding ratio $\mathcal{R} = r_\mathrm{h,max}/r_\mathrm{h}$, where $r_\mathrm{h,max}$ is the comet's own most distant observed heliocentric distance. The black points represent the $r$ and $g$ magnitudes, where the $g$ magnitudes are shifted by the median $g-r$ color reported for each comet in Table~\ref{tab:look_sample}. The colored lines show the best-fit for the $Hn$, $Hab$, Fixed-$a$, Fixed-$b$, Fixed-$ab$, and Correlated-$ab$ models, determined by the least-squares method. The lower panel shows the residuals for each model. The complete figure set (23 images) is available in the online journal.}
\label{fig:figsetexample}
\end{figure*}

A summary of the parameters found for each comet for each model using a least-squares fit is shown in Table~\ref{tab:model_fits} (Appendix~\ref{app:fit_params}). Figure \ref{fig:figsetexample} shows an example of a single comet lightcurve with all model fits and the complete figure set (23 images) is available in the online journal.

\section{Methods}
\label{sec:methods}
\subsection{Goodness of Fit}
\label{sec:method_bestfit}

To identify which model best reproduces each comet's full lightcurve, we applied the $Hab$ model, its variants, and the traditional model ($Hn$) to the complete dataset for that object with least-squares and no statistical outlier rejection, and evaluated performance with the root mean squared error (RMSE). This metric is the square root of the average squared difference between the observed and predicted magnitudes, providing an interpretable measure of model accuracy in the same units as the data. The residual sum of squares (RSS), which forms the basis of the RMSE calculation, is defined as:

\begin{equation}
\mathrm{RSS} = \sum_{i=1}^{N} \left( m_{\mathrm{obs},i} - m_{\mathrm{model},i} \right)^2
\label{eqn:rss}
\end{equation}

\noindent where $ m_{\mathrm{obs},i} $ and $ m_{\mathrm{model},i} $ are the observed and model-predicted magnitudes for each of the $N$ observations. The RMSE is then given by:

\begin{equation}
\mathrm{RMSE} = \sqrt{ \frac{\mathrm{RSS}}{N} }
\label{eqn:rmse}
\end{equation}

\noindent Lower RMSE values indicate a closer match between the model and the observed lightcurve. Photometric uncertainties were not incorporated into any of the RMSE calculations in this work, as they were generally at the $\sim$10~mmag level and did not significantly affect the results. By comparing RMSE values across models, we determined which model most accurately captured the overall brightness evolution for each comet. We note that frequent outbursts could bias this value with larger average residuals even with a generally good fit to the quiescent  periods of activity. However, none of the comets in our sample had clear outbursts beyond $r_\mathrm{h}=2.5$~au. We discuss outbursts, along with other possible causes of deviations of measurements from model fits, in Section \ref{sec:discussion}.

While a lower RMSE indicates a better fit, it does not account for the fact that a model with more free parameters will naturally tend to fit the data more closely. To evaluate whether the added complexity of the $Hab$ model was warranted relative to the traditional model ($Hn$) and its own constrained variants, we compared their fits using the Bayesian Information Criterion \citep[BIC;][]{schwarz78}, which penalizes models for additional free parameters while rewarding goodness of fit. BIC is defined as:
\vspace{-6pt}
\begin{equation}
\mathrm{BIC} = k \ln{(N)} + N \ln{\left(\frac{\mathrm{RSS}}{N}\right)}
\label{eqn:bic}
\end{equation}

\noindent where $k$ is the number of free model parameters and RSS is the residual sum of squares defined above. This formulation penalizes additional parameters more strongly than other criteria and is particularly well-suited for identifying the most appropriate model when the goal is to characterize underlying physical behavior \citep{burnham04}. \rev{For nested models such as these (each is a constrained form of $Hab$), $\Delta$BIC is equivalent to the Likelihood Ratio Test (LRT) statistic with an additional penalty of $\Delta k \ln{(N)}$. We adopted BIC because, unlike an LRT significance test, it can indicate support for the simpler model as well as the more complex one.} For each object, we considered a model to be significantly favored if its BIC value was more than 10 lower than that of a simpler model, a threshold commonly interpreted as strong evidence in model selection \citep{raftery95}\rev{; with the $\Delta k \ln(N)$ term, this requires a considerably larger improvement in fit than a conventional LRT significance test}.

\subsection{Predictive Tests}
\label{sec:method_predict}

\begin{figure*}[ht]
\begin{center}
  \includegraphics[width=.49\textwidth]{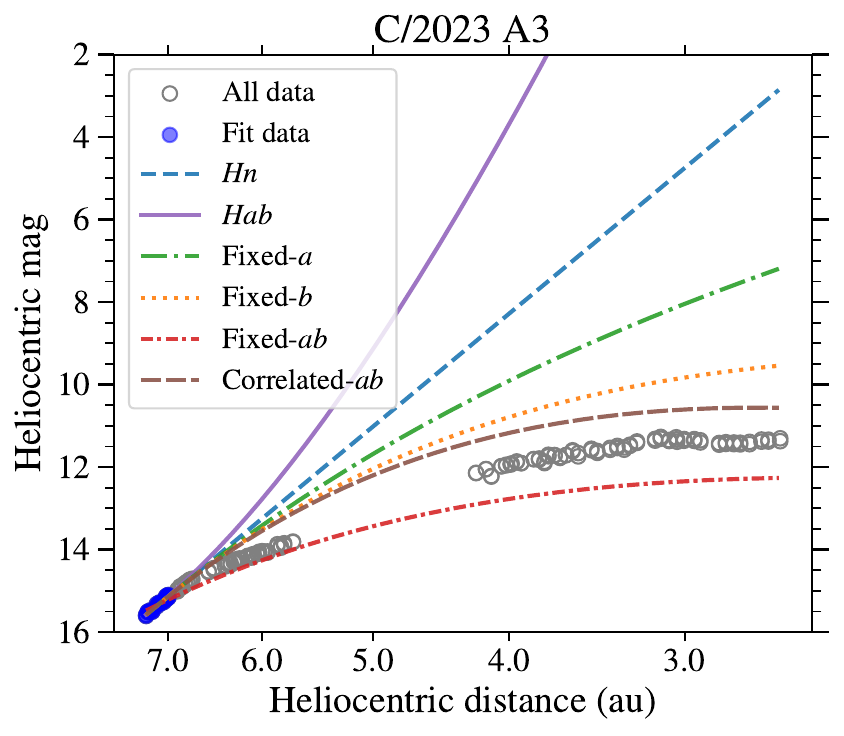}
  \includegraphics[width=.49\textwidth]{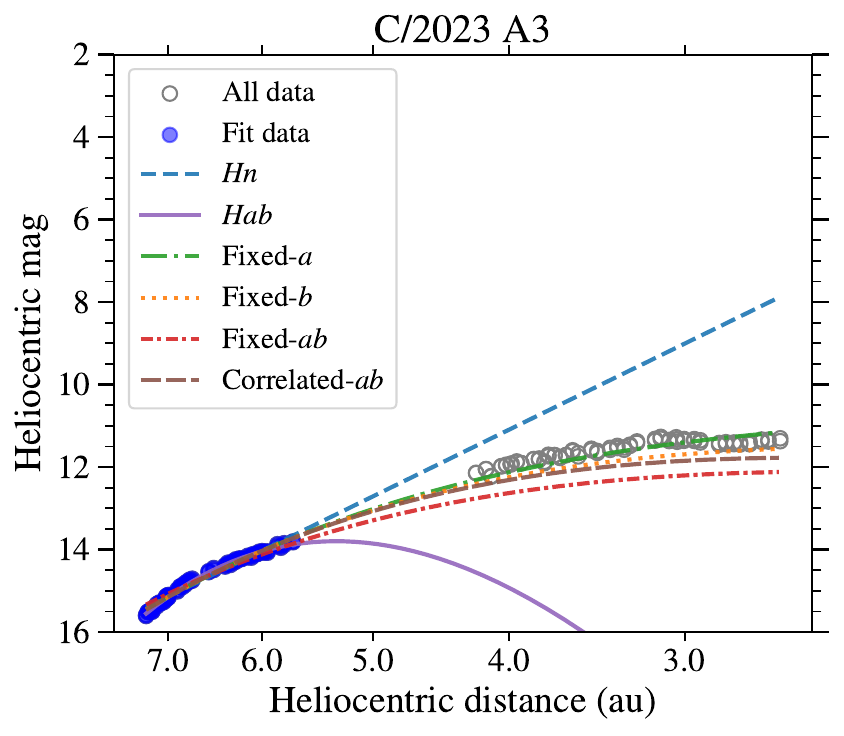}
\end{center}
\caption{An example of the strategy used for assessing predictive power of each of the models, for C/2023 A3 ($r_\mathrm{h,max} = 7.25$~au). The blue circles are the real data points used for each model fitting; the gray open circles are the rest of the real data. Each model is fit to the blue points and extrapolated out to the end of the pre-perihelion data, represented by the colored lines. The left panel uses $\mathcal{R}_\mathrm{fit} = 1.05$ ($r_\mathrm{h,fit,min} = 6.90$~au) and the right panel uses $\mathcal{R}_\mathrm{fit} = 1.3$ ($r_\mathrm{h,fit,min} = 5.58$~au).}
\label{fig:predictexample}
\end{figure*}

To evaluate the predictive performance of the $Hab$ model and its constrained variants relative to the $Hn$ model, we conducted a progressive fitting analysis by incrementally increasing the data used for each least-squares fit (i.e., decreasing $r_\mathrm{h,fit,min}$) in steps of $\Delta\mathcal{R}_\mathrm{fit} = 0.05$ (Figure~\ref{fig:predictexample}). We use $\mathcal{R}_\mathrm{fit}$ rather than a fixed arc length $\Delta r_\mathrm{h,fit}$ as the threshold variable because our models operate in $\log_{10} r_\mathrm{h}$ (Section~\ref{sec:brightening_models}); a fixed ratio corresponds to a fixed span in log-space and therefore comparable leverage on the fit at any heliocentric distance, whereas a fixed $\Delta r_\mathrm{h,fit}$ would represent very different amounts of log-space coverage depending on where along the arc it falls. For example, for C/2021~S3 ($r_\mathrm{h,max} = 8.58$~au), the $\mathcal{R}_\mathrm{fit} = 1.5$ threshold corresponds to $r_\mathrm{h,fit,min} = 5.72$~au, i.e., a fit arc of $\Delta r_\mathrm{h,fit} = 2.86$~au; for C/2023 C2 ($r_\mathrm{h,max} = 5.97$~au), the same $\mathcal{R}_\mathrm{fit} = 1.5$ threshold corresponds to $r_\mathrm{h,fit,min} = 3.98$~au, or $\Delta r_\mathrm{h,fit} = 1.99$~au. At each step, the model was fit using only the data within the specified range, and we computed the root mean squared error against the observed data up to a common $\mathcal{R} = 2$ endpoint for every comet ($\mathrm{RMSE}_{\mathcal{R}=2}$, including both the fitted and predictive portions within that range), so that predictive accuracy is compared over an equivalent relative arc across all seven comets rather than each comet's full, unevenly extended dataset. We also calculated the average magnitude difference between the predicted and observed values for the five data points nearest to, but not short of, this same $\mathcal{R} = 2$ endpoint ($|\Delta\mathrm{mag}|$), providing a complementary measure of end-of-curve predictive performance at a single reference point rather than averaged over the capped arc. We consider a prediction successful once RMSE$_{\mathcal{R}=2}$ falls below 0.5 mag or $|\Delta \mathrm{mag}|$ falls below 1 mag. RMSE$_{\mathcal{R}=2}$ gets the tighter threshold because it averages over the whole curve, while $|\Delta \mathrm{mag}|$ comes from just five points.

\rev{We initially interpolated the lightcurves onto evenly spaced points to control for uneven data sampling, but found no statistically significant difference in predictive performance between the real and interpolated datasets for any model except $Hab$; for $Hab$, the real data produced higher residuals, indicating that interpolation would understate its predictive limitations. We therefore report results from the real data throughout.}

\section{Results}
\label{sec:results}

\subsection{Best-Fit Model}
\label{sec:results_bestfit}

Using the seven comets with $\mathcal{R} > 2$, we found that the $Hab$ model provides the best fit to the full lightcurve in all seven cases. Here, each model was fit once to the complete pre-perihelion dataset for a given comet (Section~\ref{sec:method_bestfit}), rather than fit progressively and evaluated on its ability to predict unseen data, which we address separately in Section~\ref{sec:results_predict}. The full ranking of all six models by RMSE for each comet is shown in Figure~\ref{fig:rmse_rank} (Appendix~\ref{app:rankings}). It is worth noting that all models, including the traditional $Hn$ model, had an RMSE (i.e., average residuals) of less than 0.4~mag when fit to the complete lightcurve. For the entire sample of 23 comets, $Hab$ was again the best full-lightcurve fit by RMSE in every case, and the full ranking breakdown is likewise shown in Figure~\ref{fig:rmse_rank}. For all models, the RMSE did not exceed 0.55~mag.

Because $Hab$ is the clear best fit, we compared its BIC to that of the other models to assess whether the additional parameters are warranted. A $\Delta\mathrm{BIC} = \mathrm{BIC}_\mathrm{model} - \mathrm{BIC}_{Hab} > 10$ would be strong evidence that the $Hab$ model is significantly favored. We first evaluated the subset of seven comets with $\mathcal{R} > 2$, finding that $Hab$ is strongly preferred over $Hn$ in six of seven objects, over Fixed-$ab$ in all seven, over Fixed-$a$ in five of seven, and over Fixed-$b$ and Correlated-$ab$ in four of seven. Extending to the full sample of 23 comets, the $Hab$ model strongly outperforms all alternatives, with $\Delta\mathrm{BIC} > 10$ in 16-23 of 23 objects ($\gtrsim$70\% of the sample) across all pairwise comparisons. Correlated-$ab$ is the closest competitor to $Hab$ across the full sample: $Hab$ is strongly preferred ($\Delta\mathrm{BIC} > 10$) in 16 of 23 comets, weakly preferred in 5, and disfavored relative to Correlated-$ab$ in the remaining 2 (C/2021 S3, C/2021 G2). The $\Delta$BIC values for each comet and model are shown in Figure~\ref{fig:delta_bic} (Appendix~\ref{app:rankings}). Table~\ref{tab:model_fit_stats} summarizes the output of the statistical tests for each comet and model.

\begin{deluxetable*}{l r r c c r r c r r c c r r c c r r c c r r c}
\tablecaption{Model-fit statistics summary\label{tab:model_fit_stats}}
\tablecolumns{23}
\tablewidth{0pt}
\tabletypesize{\scriptsize}
\setlength{\tabcolsep}{0.04in}
\renewcommand{\arraystretch}{0.9}
\tablehead{
  \colhead{Comet} & \multicolumn{3}{c}{$\mathit{Hn}$} && \multicolumn{2}{c}{$\mathit{Hab}$} && \multicolumn{3}{c}{Fixed-$\mathit{ab}$} && \multicolumn{3}{c}{Fixed-$\mathit{a}$} && \multicolumn{3}{c}{Fixed-$\mathit{b}$} && \multicolumn{3}{c}{Correlated-$\mathit{ab}$} \\
  \cline{2-4}\cline{6-7}\cline{9-11}\cline{13-15}\cline{17-19}\cline{21-23}
  \colhead{Name} & \colhead{RSS} & \colhead{RMSE} & \colhead{$\Delta$BIC} && \colhead{RSS} & \colhead{RMSE} && \colhead{RSS} & \colhead{RMSE} & \colhead{$\Delta$BIC} && \colhead{RSS} & \colhead{RMSE} & \colhead{$\Delta$BIC} && \colhead{RSS} & \colhead{RMSE} & \colhead{$\Delta$BIC} && \colhead{RSS} & \colhead{RMSE} & \colhead{$\Delta$BIC}
}
\startdata
C/2023 A3 & 19.49 & 0.35 & 446 &  & 1.20 & 0.09 &  & 24.71 & 0.39 & 480 &  & 2.05 & 0.11 & 81 &  & 1.25 & 0.09 & 2 &  & 1.48 & 0.10 & 29 \\
C/2020 R7 & 5.35 & 0.17 & 245 &  & 1.38 & 0.09 &  & 2.44 & 0.11 & 95 &  & 2.16 & 0.11 & 77 &  & 1.95 & 0.10 & 59 &  & 1.66 & 0.09 & 29 \\
C/2021 T4 & 3.82 & 0.26 & 167 &  & 0.20 & 0.06 &  & 3.22 & 0.24 & 153 &  & 0.32 & 0.07 & 23 &  & 0.21 & 0.06 & -1 &  & 0.22 & 0.06 & 3 \\
C/2022 L2 & 4.77 & 0.20 & 290 &  & 0.38 & 0.06 &  & 0.59 & 0.07 & 41 &  & 0.39 & 0.06 & -4 &  & 0.38 & 0.06 & -5 &  & 0.41 & 0.06 & 2 \\
C/2022 E2 & 9.03 & 0.22 & 378 &  & 1.06 & 0.08 &  & 4.12 & 0.15 & 233 &  & 1.10 & 0.08 & 1 &  & 1.26 & 0.08 & 25 &  & 1.46 & 0.09 & 52 \\
C/2021 S3 & 4.23 & 0.17 & 426 &  & 0.20 & 0.04 &  & 7.91 & 0.24 & 508 &  & 0.55 & 0.06 & 140 &  & 0.25 & 0.04 & 27 &  & 0.20 & 0.04 & -5 \\
C/2023 C2 & 0.38 & 0.10 & 0 &  & 0.35 & 0.10 &  & 5.56 & 0.38 & 98 &  & 0.92 & 0.16 & 34 &  & 1.17 & 0.18 & 43 &  & 1.64 & 0.21 & 55 \\
\hline
C/2022 S4 & 11.40 & 0.30 & 259 &  & 1.40 & 0.11 &  & 8.31 & 0.26 & 215 &  & 4.42 & 0.19 & 140 &  & 3.71 & 0.17 & 118 &  & 2.98 & 0.15 & 90 \\
C/2021 G2 & 3.01 & 0.12 & 233 &  & 0.96 & 0.07 &  & 56.89 & 0.52 & 839 &  & 2.66 & 0.11 & 207 &  & 1.18 & 0.08 & 38 &  & 0.96 & 0.07 & -4 \\
C/2024 E1 & 0.55 & 0.09 & -2 &  & 0.53 & 0.09 &  & 2.95 & 0.21 & 101 &  & 1.55 & 0.16 & 64 &  & 1.34 & 0.14 & 55 &  & 1.21 & 0.14 & 48 \\
C/2019 L3 & 0.16 & 0.05 & -4 &  & 0.16 & 0.05 &  & 22.71 & 0.54 & 374 &  & 0.28 & 0.06 & 40 &  & 0.50 & 0.08 & 84 &  & 0.96 & 0.11 & 135 \\
C/2022 QE78 & 2.19 & 0.20 & 170 &  & 0.10 & 0.04 &  & 9.36 & 0.41 & 249 &  & 0.43 & 0.09 & 77 &  & 0.73 & 0.11 & 107 &  & 0.88 & 0.12 & 118 \\
C/2022 A3 & 1.42 & 0.12 & 129 &  & 0.36 & 0.06 &  & 3.76 & 0.20 & 220 &  & 0.65 & 0.08 & 52 &  & 0.50 & 0.07 & 26 &  & 0.39 & 0.06 & 3 \\
C/2023 R1 & 1.24 & 0.12 & 134 &  & 0.26 & 0.05 &  & 3.38 & 0.20 & 218 &  & 0.38 & 0.07 & 31 &  & 0.47 & 0.07 & 50 &  & 0.53 & 0.08 & 59 \\
C/2024 A1 & 0.87 & 0.10 & 0 &  & 0.83 & 0.10 &  & 1.14 & 0.11 & 19 &  & 1.12 & 0.11 & 21 &  & 1.12 & 0.11 & 22 &  & 1.13 & 0.11 & 22 \\
C/2022 E3 & 0.09 & 0.06 & 21 &  & 0.03 & 0.04 &  & 3.62 & 0.39 & 105 &  & 0.04 & 0.04 & 3 &  & 0.04 & 0.04 & -2 &  & 2.60 & 0.33 & 100 \\
C/2023 T3 & 2.31 & 0.21 & 108 &  & 0.28 & 0.07 &  & 2.35 & 0.21 & 105 &  & 1.58 & 0.17 & 88 &  & 1.46 & 0.17 & 84 &  & 1.25 & 0.15 & 76 \\
C/2021 E3 & 0.11 & 0.05 & 26 &  & 0.05 & 0.04 &  & 3.59 & 0.31 & 151 &  & 0.05 & 0.04 & -1 &  & 0.05 & 0.04 & 0 &  & 0.24 & 0.08 & 55 \\
C/2022 R6 & 3.80 & 0.15 & 313 &  & 0.63 & 0.06 &  & 6.39 & 0.19 & 399 &  & 1.53 & 0.09 & 151 &  & 1.84 & 0.10 & 184 &  & 1.96 & 0.11 & 195 \\
C/2021 O3 & 0.19 & 0.09 & 54 &  & 0.02 & 0.03 &  & 2.24 & 0.31 & 110 &  & 0.11 & 0.07 & 41 &  & 0.06 & 0.05 & 27 &  & 0.03 & 0.04 & 10 \\
C/2022 U3 & 0.97 & 0.14 & 45 &  & 0.39 & 0.09 &  & 1.19 & 0.15 & 52 &  & 0.77 & 0.12 & 33 &  & 0.75 & 0.12 & 31 &  & 0.74 & 0.12 & 30 \\
C/2022 T1 & 0.43 & 0.10 & 94 &  & 0.04 & 0.03 &  & 0.63 & 0.12 & 106 &  & 0.27 & 0.08 & 74 &  & 0.30 & 0.08 & 79 &  & 0.37 & 0.09 & 87 \\
C/2021 C5 & 0.11 & 0.05 & 0 &  & 0.10 & 0.05 &  & 1.33 & 0.17 & 111 &  & 0.10 & 0.05 & -3 &  & 0.11 & 0.05 & 1 &  & 0.13 & 0.05 & 9 \\
\enddata
\begin{tablenotes}[para,flushleft]
\item RSS: The residual sum of squares for each model and each comet (Equation \ref{eqn:rss}) in units of mag$^2$. The RSS forms the basis of the RMSE equation. A smaller value indicates a better fit. \\
\item RMSE: The root mean squared error based on the RSS for each model and each comet (Equation \ref{eqn:rmse}) in units of mag. A smaller value indicates a better fit. \\
\item $\Delta\mathrm{BIC}$: $\mathrm{BIC}_{\mathrm{model}} - \mathrm{BIC}_{Hab}$. Positive values indicate the $Hab$ model is preferred; $\Delta\mathrm{BIC} > 10$ indicates strong preference. Negative values indicate the alternative model is preferred over $Hab$. The $Hab$ $\Delta\mathrm{BIC}$ column is omitted since $\Delta\mathrm{BIC} = 0$ by definition. \\
\item The horizontal line separates the seven comets with the largest $\mathcal{R}$ (above) from the rest of the sample (below)
\end{tablenotes}
\end{deluxetable*}

\vspace{-20pt}
\subsection{Predictive Power}
\label{sec:results_predict}

\begin{figure}[ht]
    \centering
    \includegraphics[width=\columnwidth]{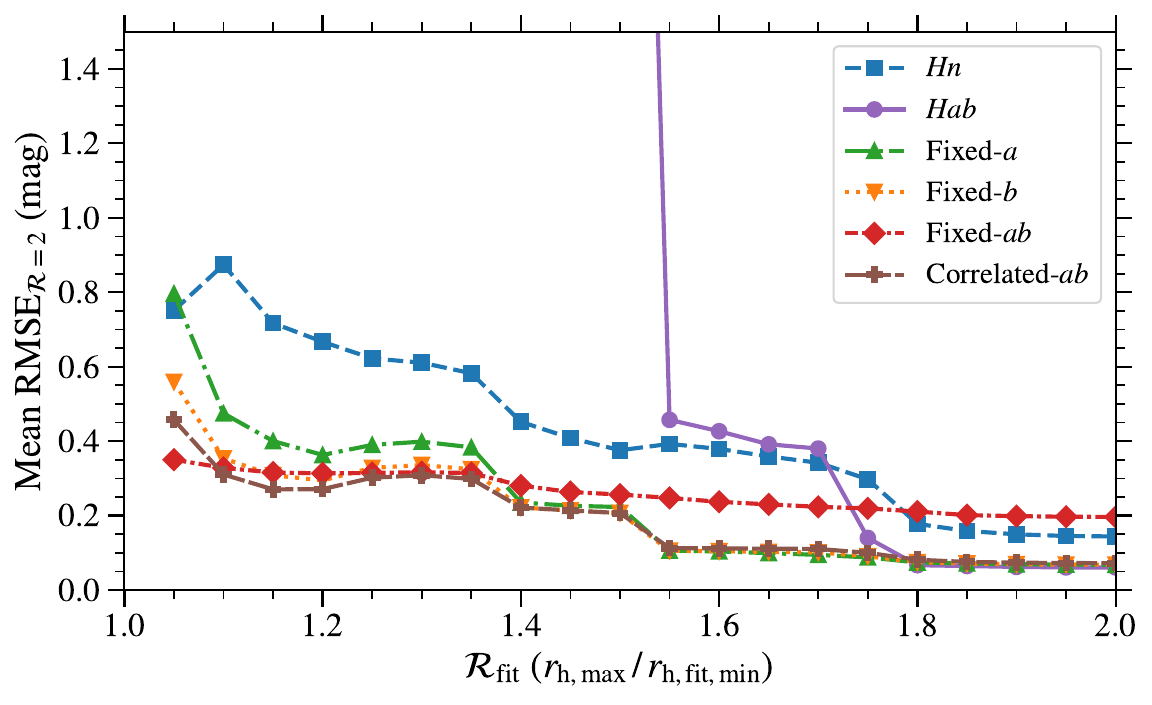}
    \caption{Mean root-mean-squared error (RMSE$_{\mathcal{R}=2}$) up to the common $\mathcal{R}=2$ endpoint, as a function of the ratio threshold $\mathcal{R}_\mathrm{fit}$ used for fitting, averaged across the seven $\mathcal{R} > 2$ comets. The curves are not perfectly smooth, reflecting the small number of comets (N = 7) contributing to the mean, each sampled at its own irregular real cadence (see Appendix B, Figures \ref{fig:rmse_permodel} and \ref{fig:R_vs_deltarh}). Fixed-$ab$ and Correlated-$ab$ reach RMSE$_{\mathcal{R}=2}$ below 1 mag for all seven comets from the earliest threshold tested, while $Hab$ and $Hn$ require substantially more data to converge. At large $\mathcal{R}_\mathrm{fit}$, $Hab$ and Correlated-$ab$ converge to comparable RMSE$_{\mathcal{R}=2}$; note that this predictive comparison uses only data up to a given $\mathcal{R}_\mathrm{fit}$ threshold, capped at $\mathcal{R} = 2$, whereas the full-lightcurve comparison in Section~\ref{sec:results_bestfit} fits and evaluates each model using the complete dataset for each comet (extending to each comet's full observed $\mathcal{R}$); the two are therefore not evaluating the models against the same amount of data.}
    \label{fig:rmse_predict}
\end{figure}

\begin{figure}[ht]
    \centering
    \includegraphics[width=\columnwidth]{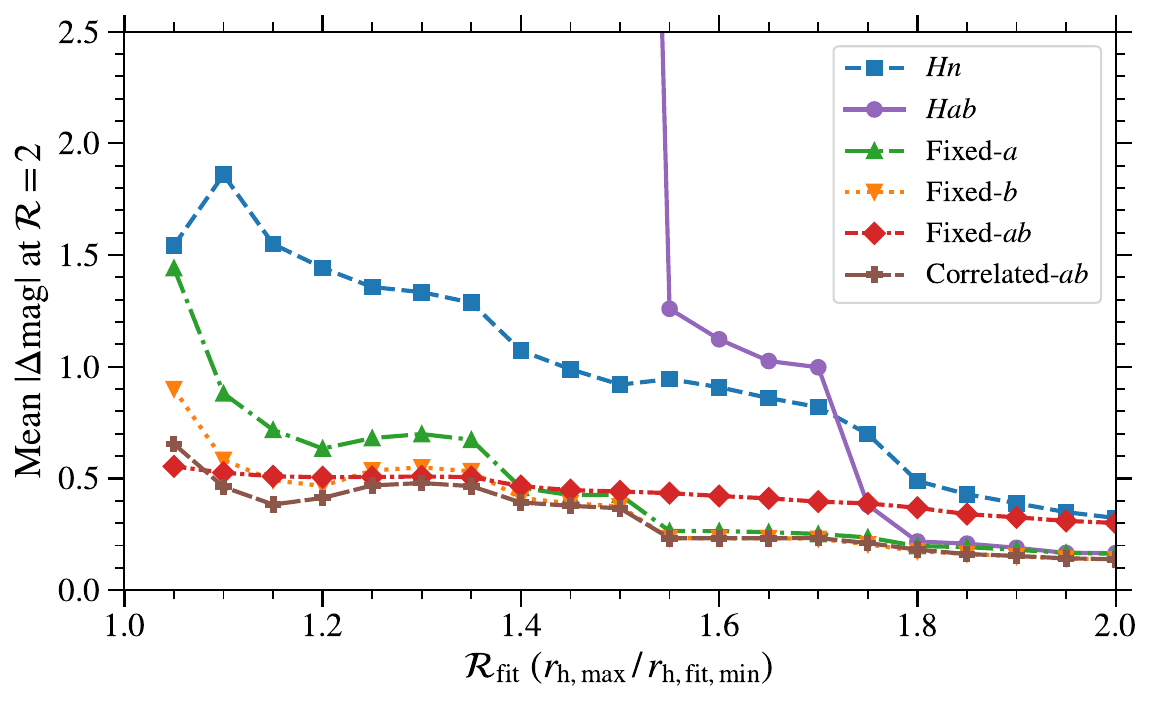}
    \caption{Mean absolute magnitude difference between each model's prediction and the observed brightness at a fixed evaluation point, $\mathcal{R} \approx 2$, as a function of the ratio threshold $\mathcal{R}_\mathrm{fit}$ used for fitting. This metric isolates predictive accuracy at a single distant reference point rather than averaging over the full extrapolated curve as in Figure \ref{fig:rmse_predict}. It shows a broadly consistent picture: Fixed-$ab$ and Correlated-$ab$ converge earliest, while $Hab$ and $Hn$ remain unreliable until $\mathcal{R}_\mathrm{fit} \gtrsim 1.8$.}
    \label{fig:mag_predict}
\end{figure}

\rev{Figures \ref{fig:rmse_predict} and \ref{fig:mag_predict} show the mean RMSE$_{\mathcal{R}=2}$ and mean $|\Delta \mathrm{mag}|$ at $\mathcal{R} = 2$ (Section \ref{sec:method_predict}), respectively, across our seven $\mathcal{R} > 2$ comets as a function of $\mathcal{R}_\mathrm{fit}$.} With only seven comets contributing to the mean, each sampled at its own irregular real cadence with gaps during solar conjunction, both aggregate curves show modest non-monotonic structure around $\mathcal{R}_\mathrm{fit} \sim 1.2$ -- 1.4 rather than a perfectly smooth trend. Appendix \ref{app:predict_diagnostics} shows this is a small-sample sampling effect rather than model instability: the individual per-comet RMSE$_{\mathcal{R}=2}$ curves (Figure \ref{fig:rmse_permodel}) and the corresponding fit-arc length vs. $\mathcal{R}_\mathrm{fit}$ (Figure \ref{fig:R_vs_deltarh}) show that the fit-arc used at a given $\mathcal{R}_\mathrm{fit}$ changes in discrete steps as the threshold sweep crosses real observation epochs, not continuously.

$Hab$ is the best-fit model for characterizing the full lightcurve (Section \ref{sec:results_bestfit}) but the worst-performing for prediction from limited early data: at $\mathcal{R}_\mathrm{fit} \lesssim 1.3$ ($\Delta r_\mathrm{h,fit} \sim 1.5$ -- 2 au), $Hab$ fails badly by both measures (mean $|\Delta \mathrm{mag}|$ exceeds 10 mag throughout; mean RMSE$_{\mathcal{R}=2}$ exceeds 10 mag at $\mathcal{R}_\mathrm{fit} = 1.05$ and stays high, $\sim$7 mag, through 1.20), reflecting the difficulty of constraining three free parameters from a short arc. $Hn$ performs similarly poorly at small $\mathcal{R}_\mathrm{fit}$, succeeding for only 2 -- 4 of 7 comets by either metric until $\mathcal{R}_\mathrm{fit} \sim 1.4$.

Fixed-$ab$ is the only model to predict brightnesses for all 7 comets to within 1 mag at the very first threshold tested ($\mathcal{R}_\mathrm{fit} = 1.05$, mean RMSE$_{\mathcal{R}=2}$ $=$ 0.35 mag, mean $|\Delta \mathrm{mag}| = 0.55$ mag), since fixing both $a$ and $b$ leaves only $H(1,1,0)$ free: with $a$ and $b$ fixed, $H$ sets only the vertical offset of the curve rather than its shape, so it can be constrained from even a single data point. This advantage is short-lived, however: by the next threshold ($\mathcal{R}_\mathrm{fit} = 1.10$), Correlated-$ab$ is already more accurate than Fixed-$ab$ by both metrics (RMSE$_{\mathcal{R}=2}$ $=$ 0.31 vs. 0.33 mag; $|\Delta \mathrm{mag}|$ $=$ 0.46 vs. 0.53 mag), and the gap widens steadily thereafter, reaching roughly a third of Fixed-$ab$'s error by $\mathcal{R}_\mathrm{fit} \sim 2$ (RMSE$_{\mathcal{R}=2} =$ 0.07 vs. 0.20 mag). Fixed-$ab$'s value is therefore specifically in guaranteeing a usable prediction from the very first available epoch, not in being the most accurate choice once even slightly more data exist. Fixed-$b$ succeeds for only 2/7 comets at $\mathcal{R}_\mathrm{fit} = 1.05$ by both metrics, catching up to Correlated-$ab$ by $\mathcal{R}_\mathrm{fit} \sim 1.55$. Fixed-$a$ is the slowest of the four constrained variants under both metrics, not reaching full success until $\mathcal{R}_\mathrm{fit} \sim 1.55$.

All models converge toward good predictive performance by both metrics once enough data are available ($\mathcal{R}_\mathrm{fit} \gtrsim 1.8$), consistent with $Hab$'s superior full-curve characterization once sufficiently constrained.

\subsection{Characterization}
\label{sec:results_characterization}

\begin{figure*}[ht]
    \begin{center}
      \includegraphics[width=.49\textwidth]{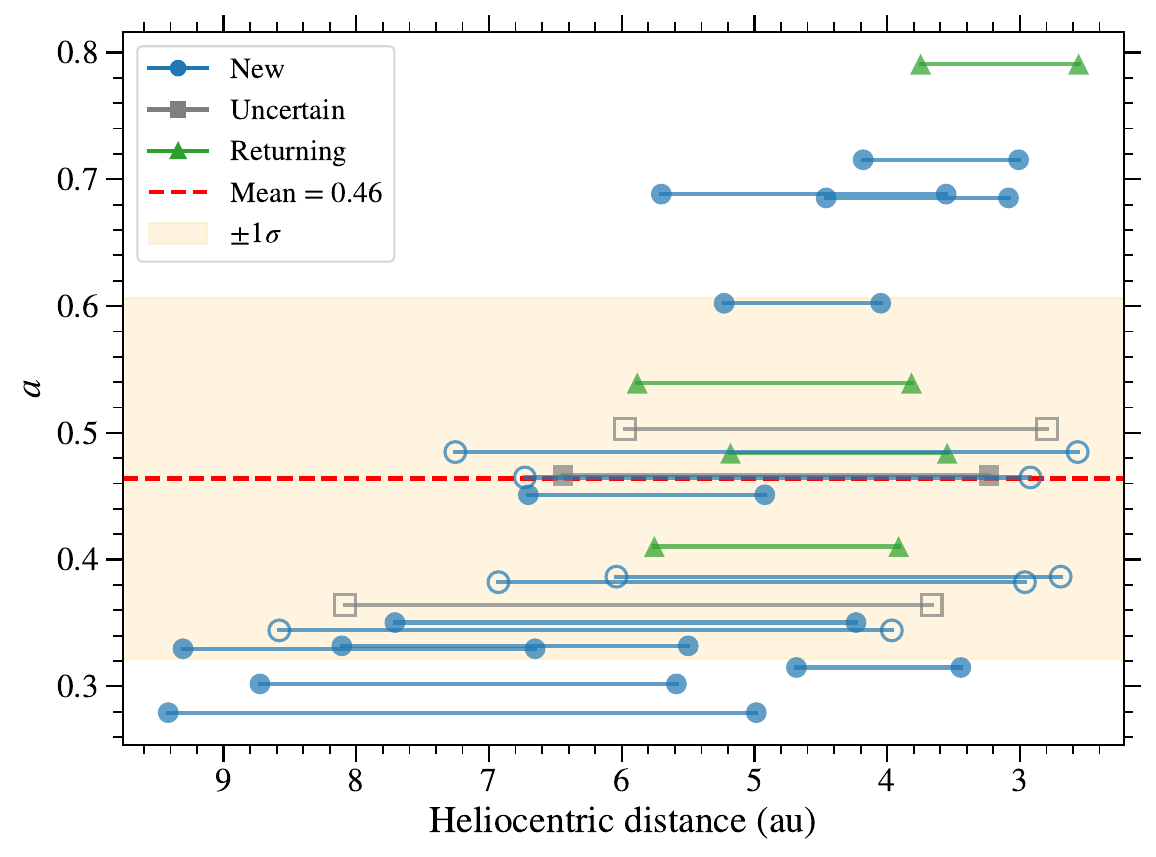}
      \includegraphics[width=.49\textwidth]{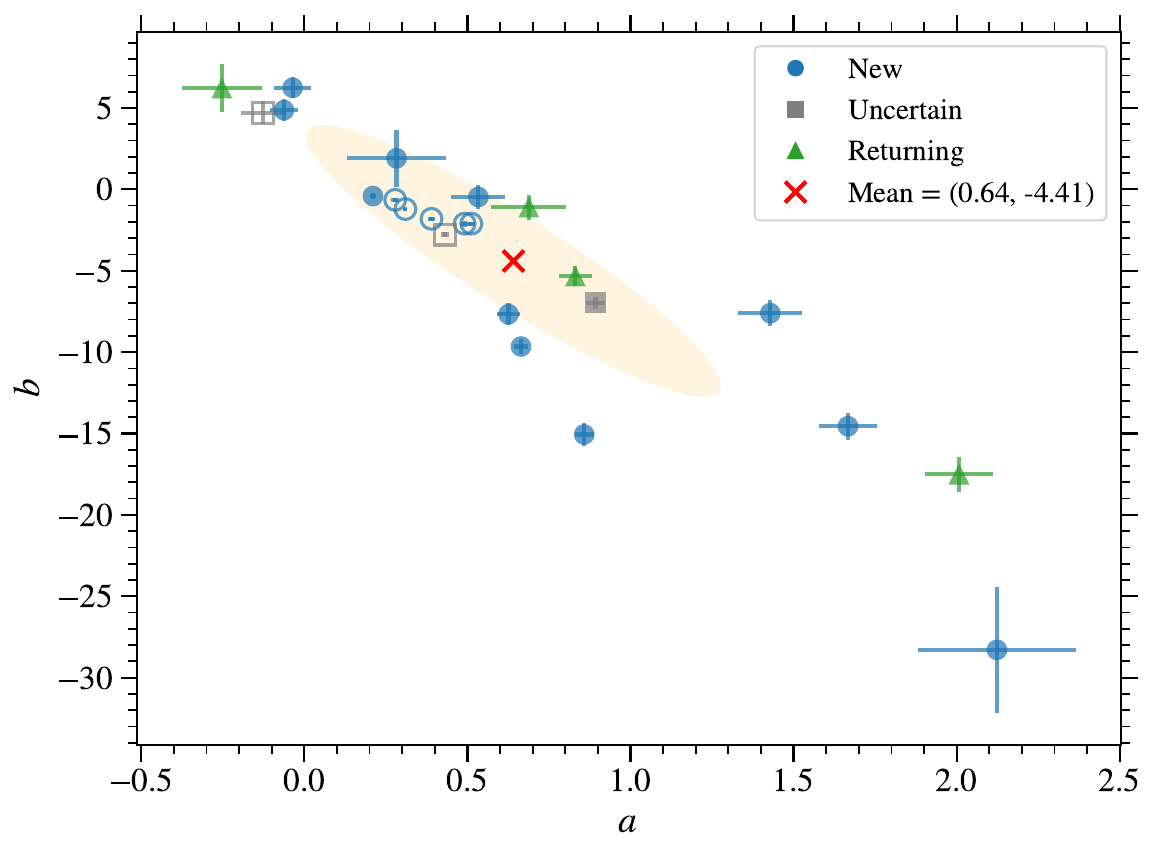}
    \end{center}
    \caption{
    (left) Heliocentric distance of the data for each comet, plotted against the coefficient $a$ from the Fixed-$b$ model fit. Comets are colored and shaped by dynamical class inferred from $1/a_0$: New ($1/a_0 < 40\times10^{-6}$~au$^{-1}$; blue circles), Uncertain ($40$ -- $100\times10^{-6}$~au$^{-1}$; gray squares), and Returning ($>100\times10^{-6}$~au$^{-1}$; green triangles). Open symbols mark the seven comets with $\mathcal{R} > 2$ used in our high quality tests. The dashed red line and shaded band show the sample mean ($a = 0.46$) and $\pm1\sigma$ range. Comets sampled closer to the Sun tend toward systematically larger $a$, and these are also the comets with some of the smallest $\Delta r_\mathrm{h}$ coverage. (right) The $Hab$-model parameters $a$ and $b$ for the same sample, colored and marked as in the left panel; open symbols again denote the seven $\mathcal{R} > 2$ comets. The red cross and shaded ellipse show the mean and $1\sigma$ covariance region for the \textit{entire} sample. For both models, we observe no trend with dynamical class. 
    \label{fig:a_dynclass}
    }
\end{figure*}

After assessing the distribution of the $a$ and $b$ parameters in the $Hab$ model (Section~\ref{sec:3param}, Figure~\ref{fig:threeparamdist}), we examined the distribution of $a$ in the two-parameter models. We focused on the Fixed-$b$ and Correlated-$ab$ models, since Fixed-$a$ was the worst-performing two-parameter variant in our predictive tests (Section~\ref{sec:results_predict}) and is not discussed further here. Fixed-$b$ was the second-best fit for three of the seven high-quality ($\revmath{\mathcal{R}>2}$) comets (Section~\ref{sec:results_bestfit}). For this model, $a$ has a mean of 0.46 (median of 0.45) with a $1\sigma$ standard deviation of 0.14 across the full sample; for comparison, the Correlated-$ab$ model gives a similar but slightly broader distribution, with a mean $a$ of 0.44 (median of 0.36) and a $1\sigma$ standard deviation of 0.18, after $3\sigma$ outlier rejection excluded 4 of the 23 comets. We also found that larger $a$ values correspond to comets sampled over smaller heliocentric distance ranges closer to the Sun (Figure~\ref{fig:a_dynclass}, left panel).

We further examined whether either model's parameters correlate with dynamical class. Comets with $1/a_0 < 4\times10^{-5}$~au$^{-1}$ are classified as New, following the threshold adopted in Section~\ref{sec:intro_limitations}, while comets with $1/a_0 > 10^{-4}$~au$^{-1}$, the New threshold used by \citet{lacerda25} for the same purpose, are classified as Returning. Comets falling between these two literature thresholds are classified as Uncertain. As shown in Figure~\ref{fig:a_dynclass}, we found no clear trend with dynamical class in the $a$ values from the Fixed-$b$ model or in the full-sample $Hab$ $(a,b)$ distribution (mean $a = 0.64$, $b = -4.41$; median $a = 0.51$, $b = -2.11$), consistent with the lack of a dynamical-class dependence found by \citet{holt24}.

\break
\section{Discussion}
\label{sec:discussion}

\subsection{An Improved Brightening Model}
\label{sec:improved_model}

Allowing the activity index to vary linearly with $r_\mathrm{h}$ greatly improves the fits to the observations. This gives smoothly varying heliocentric lightcurves that fit all comets in our sample remarkably well. It is also notable that the range of fit parameters for the best sampled comets (those observed over the largest range in \rh) is small, and that even the Fixed-$ab$ model does a good job of approximating the behavior of most of the comets. This suggests a common activity pattern across most long-period comets discovered around $r_\mathrm{h}=5-9$~au, which we can describe by looking at the $n_\mathrm{eff}$ with \rh\  for the average case (i.e., the Fixed-$ab$ model; Figure \ref{fig:holt24fig}). This has $n_\mathrm{eff} \sim 2$ around \rh\ = 4 au, corresponding to the `typical' value for historically observed DNCs, where the amount of material in the coma is constant, and the change in brightness is just due to the inverse-square law illumination change. This makes sense, as most of the comets in our sample are DNCs, although we found no difference in brightening behavior between DNCs and returning comets in this sample (Section~\ref{sec:results_characterization}). The Fixed-$ab$ model has $n_\mathrm{eff}$ reach zero at \rh\ = 2.3 au (and go negative inside this distance), showing the limits of this empirical model -- it fits well for the distant evolution of comet brightness, but not in the water-dominated region, so should not be relied on to describe activity changes close to the Sun (e.g., $r_\mathrm{h} \lesssim 2.5$~au). At larger distances the average comet described by the Fixed-$ab$ model has a rapid increase in brightness, with $n_\mathrm{eff} \sim 10$ at \rh\ = 10 au and $\sim 20$ at 15 au, suggesting activity `turning on' at these distances. This appears to be corroborated by the observed range of discovery distances of comets, which in the modern survey era are regularly found out to \rh\ $\sim$ 10 au, but rarely beyond that distance \citep{lister22}.

Around this general behavior, there are clearly subtleties for individual comets, and there are other comets that are clear outliers -- for example, those showing very distant activity at 20 au or more (e.g., C/2017 K2) -- which were not included in the sample used here. One interpretation of this could be that the majority of LPCs we observe are depleted in near-surface CO (perhaps due to heating before they were placed into the Oort cloud; \citealt{gkotsinas24}), and their activity therefore starts when CO$_2$ driven activity becomes significant near 10 au, while more distantly active comets are the exceptional ones that retain significant CO. This would match JWST observations showing that C/2024 E1 was CO depleted \citep{snodgrass25}; this comet was discovered at $r_\mathrm{h}=8$~au and had a pre-perihelion brightness evolution that could be described by the $Hab$ model, where \rev{$a = -0.06 \pm 0.04$ and $b = 4.86 \pm 0.69$}.

Testing this interpretation directly would require volatile abundance measurements for a larger fraction of the sample, since our own dataset only consists of broadband photometry. JWST is well suited to this: NIRSpec IFU spectroscopy can constrain the CO/CO$_2$/H$_2$O abundance ratios of individual comets at the heliocentric distances relevant here, as demonstrated for C/2024~E1 \citep{snodgrass25}. Spectroscopic follow-up of newly discovered distant comets, compared against their photometric $(a,b)$ values, would offer a direct test of whether volatile composition drives the diversity seen in Figure~\ref{fig:threeparamdist}. This test would also extend to rarer distant discoveries, which may have different volatile abundances resulting in brightening behavior that differs from our sample, potentially occupying a distinct region of $(a,b)$ space.

\subsection{Interpreting \texorpdfstring{$H(1,1,0)$}{H(1,1,0)} Across Models}
\label{sec:disc_h_interp}

While Sections~\ref{sec:results_bestfit} and \ref{sec:results_predict} focus on how well each model fits and predicts brightness, the recovered $H(1,1,0)$ values themselves carry an important caveat: they are not directly comparable across models. The $H(1,1,0)$ values recovered also depend strongly on which parameters are fixed (Table~\ref{tab:model_fits}). Adopting the Fixed-$ab$ model, for example, yields $H$ values that are fainter by $6.2 \pm 2.0$~mag on average relative to the traditional $Hn$ model, a systematic offset between models rather than a physical difference in nucleus size. This model dependence means that once $a$ and $b$ are allowed to vary, $H$ loses much of its utility as a proxy for comparing intrinsic brightness between comets, limiting its applicability to metrics such as the Bortle limit \citep{bortle91}. As with comet nuclei generally, $H$ cannot be converted to a physical size using the size-albedo formalism applied to asteroids; that said, the $H$ values from the Fixed-$ab$, Fixed-$a$, Fixed-$b$, and Correlated-$ab$ models are broadly closer to plausible nucleus sizes under this formalism than those from $Hn$, whereas the full three-parameter $Hab$ model yields a much wider, unphysical range of $H$ (2.3 to 46.8~mag) across the sample.

\subsection{Predicting Brightness Behavior}

The $Hab$ model, which best reproduces a long pre-perihelion lightcurve, is not always the safest choice for forecasting when only a short early arc exists. $Hab$, Fixed-$ab$, Fixed-$a$, Fixed-$b$, and Correlated-$ab$ all share the same functional form, $ar_\mathrm{h} + b$; they differ only in whether $a$, $b$, or both are fixed to the population values $A_1$ and $B_1$, rather than left free. With both $a$ and $b$ free, $Hab$ is preferred by BIC for characterizing the distant evolution in the majority of our sample (Section~\ref{sec:results_bestfit}), though Correlated-$ab$ achieves comparable RMSE$_{\mathcal{R}=2}$ once enough data are available (Figure~\ref{fig:rmse_predict}) and reaches that level of performance with substantially less data. $Hab$'s flexibility becomes a liability when only a short arc is available to constrain three free coefficients; in our progressive fitting experiments, it is the weakest performer at forecasting further along the inbound leg by both the RMSE$_{\mathcal{R}=2}$ and fixed-point $|\Delta\mathrm{mag}|$ metrics (Section~\ref{sec:results_predict}). The traditional $Hn$ model is not markedly better, struggling even with moderate data coverage.

Table \ref{tab:recommendations} translates these thresholds into a practical recommendation for observers deciding which model to trust at a given stage of an inbound arc. The pattern underlying it, evident in the metrics reported in Section \ref{sec:results_predict}, is that fixing more coefficients trades full-curve fit quality for predictive stability when data are limited, and that trade becomes less necessary as the arc lengthens. With only discovery-epoch data ($\mathcal{R}_\mathrm{fit} \lesssim 1.05$), Fixed-$ab$ is the only model that predicts all seven comets to within 1 mag from the first available epoch, since fixing both $a$ and $b$ to $A_1$ and $B_1$ leaves $H$ as the sole free parameter. By $\mathcal{R}_\mathrm{fit} \sim 1.2$ (roughly 1 -- 1.5 au of coverage), Correlated-$ab$ overtakes it as the preferred two-coefficient choice. Once the arc reaches roughly $\Delta r_\mathrm{h,fit} \sim 2.5$ -- 4 au ($\mathcal{R}_\mathrm{fit} \gtrsim 1.8$), enough data exist to leave both $a$ and $b$ free (in addition to $H$), and the fully unconstrained $Hab$ model both meets the 1 mag threshold and gives the best characterization of the lightcurve, at least until the comet enters the water-dominated regime ($r_\mathrm{h} \lesssim 2.5$~au). Even once fully converged, the RMSE$_{\mathcal{R}=2}$ for the Fixed-$ab$ model is roughly three times that of $Hab$. Fixed-$ab$ is not the most accurate model at any stage; it is simply the only one usable immediately.

\rev{Among the three two-parameter variants}, Correlated-$ab$ reaches full-sample convergence earliest, followed closely by Fixed-$b$, with Fixed-$a$ the slowest (Section \ref{sec:results_predict}). \rev{Because the anchor values and the $a$--$b$ relation were derived from six of the seven comets used in this test, we repeated the predictive analysis using anchors recomputed without each predicted comet (Section~\ref{sec:3param}). The results were essentially unchanged ($\lesssim$0.05~mag for Correlated-$ab$, Fixed-$a$, and Fixed-$b$), since each retains a free coefficient that absorbs small shifts in the anchors; Fixed-$ab$, with no free shape parameter, was the most affected ($\lesssim$0.1~mag). Nonetheless, all of these anchors are derived from a small sample, and Fixed-$b$, which does not depend on the $a$--$b$ correlation, may be the more reliable choice if a larger dataset reveals a weaker or different relation.}

Our predictive framework could in principle also be applied in the opposite direction: rather than forecasting a comet's future brightness, one could extrapolate backward from early inbound data to estimate its brightness at larger $r_\mathrm{h}$, before discovery. This is directly relevant to precovery searches, where archival survey data are mined for earlier detections of a newly discovered comet, since a model-based estimate of the pre-discovery brightness could indicate whether such a search is worthwhile. We have not tested backward extrapolation with this dataset, though the reduced-parameter models in Table~\ref{tab:recommendations} may be similarly well suited to this application, given their performance in forward prediction.

\begin{table*}
\centering
\caption{Recommended brightening models as a function of available data 
coverage for comets discovered at $r_\mathrm{h} \sim 6$--9~au.}
\label{tab:recommendations}
\begin{tabular}{llll}
\hline\hline
$\mathcal{R}_\mathrm{fit}$ & $\Delta r_\mathrm{h,fit}$ & Recommended model & Activity Term\textsuperscript{a}\\
\hline
$\lesssim 1.2$ & $\lesssim 1$~au & Fixed-$ab$ & $A_1 r_\mathrm{h} + B_1$\\
$\gtrsim 1.2$ & $\gtrsim 1$--1.5~au & Correlated-$ab$ & $a r_\mathrm{h} -6.385a+0.792$\\
$\gtrsim 1.8$ & $\gtrsim 2.5$--4~au & $Hab$ & $a r_\mathrm{h} + b$\\
\hline
\multicolumn{4}{l}{$Hn$ is not recommended for predictive use at any stage for $r_\mathrm{h}>2.5$~au.} \\
\multicolumn{4}{l}{\textsuperscript{a}\rev{Replaces $n$ in Equation~\ref{eqn:aper_correct}, but is not equivalent to $n$ (see $n_{\rm eff}$, Equation~\ref{eqn:effective_n}).}} \\
\hline
\end{tabular}
\end{table*}

\subsection{Implications for Comet Interceptor} 

This model can hopefully assist the {\it Comet Interceptor} mission in two ways: firstly, it can be used to look at the most general evolution of distant comet brightness, as an input for simulations for likely discovery rates with LSST (Donaldson et al., in prep), which are important to assess the likelihood of discovery of suitable targets with time, and consequently inform decisions on whether to accept a possible target or wait for a `better' one \citep{snodgrass-roma,vigren23}. Secondly, and more directly, the approximations provided by the $Hab$ model variants can assist in making more realistic predictions for the future brightness evolution of a newly discovered comet --- particularly soon after discovery, when only a short arc is available ---  to assess its suitability as a target. 

\begin{figure}[ht]
    \centering
    \includegraphics[width=\columnwidth]{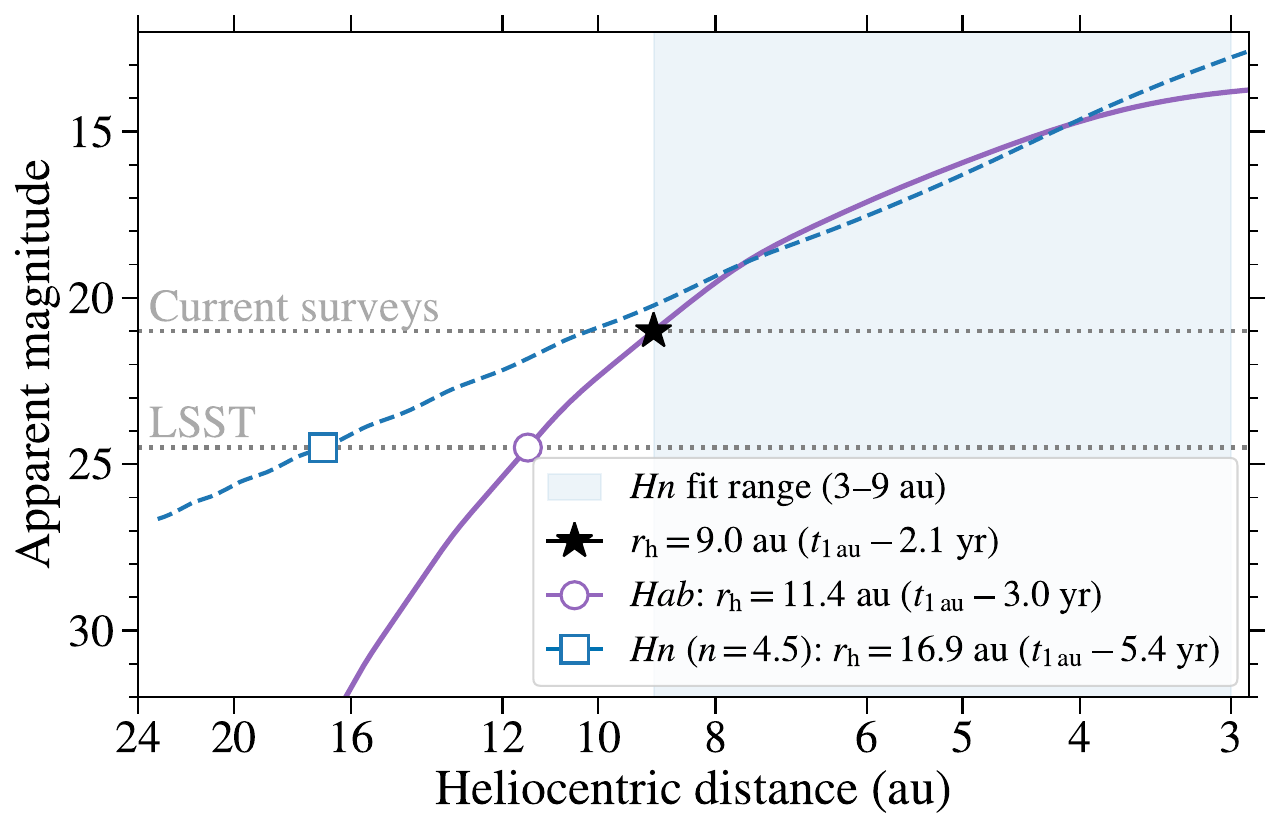}
    \caption{Apparent magnitude vs.\ heliocentric distance ($r_\mathrm{h}$, log scale) for a hypothetical comet using the real orbital geometry of a comet with $q = 0.5$~au. The comet is discovered at $r_\mathrm{h} = 9.0$~au with $m = 21.0$ (filled star), a typical current-survey limit (upper dotted line). The solid purple curve shows the $Hab$ model ($a = 0.40$, $b = -1.78$); the dashed blue curve shows the $Hn$ model \rev{fit to the $Hab$ curve over 3--9~au (shaded region; $n = 4.5$)}. Open symbols mark where each model reaches $m = 24.5$, the LSST single-visit depth (lower dotted line). Times are relative to $t_\mathrm{1\,au}$, when the comet reaches $r_\mathrm{h} = 1$~au, the approximate encounter distance of Comet Interceptor.}
    \label{fig:example_discovery_dists}
\end{figure}

It is worth noting that the finding that most comets brighten rapidly at \rh $\sim$ 10 au has important implications for the mission: if comets with activity at substantially larger distances are unusual, then even in the LSST era we may not expect to find many. This means that the likely mission target will be found closer to $r_\mathrm{h}\sim10$~au, which means only $\sim$ 3 years from perihelion for comets coming close to $r_\mathrm{h}=1$~au (Figure \ref{fig:example_discovery_dists}). \rev{By contrast, a $Hn$ model fit to the same inbound lightcurve ($n = 4.5$ over 3--9~au, consistent with the mean $n = 4.2 \pm 1.9$ for dynamically new comets from \citealt{holt24}) predicts LSST detection at $r_\mathrm{h} = 16.9$~au, $\sim$2.4~yr earlier than $Hab$. Because $n_\mathrm{eff}$ steepens beyond $\sim$10~au (Section~\ref{sec:improved_model}), any extrapolation of the traditional $Hn$ model will overestimate how early distant comets can be found.} Rapid decision making and reaction will therefore likely be necessary to meet an incoming comet with relatively little warning, given typical cruise times of 2-3 years \citep{sanchez21}.

\subsection{Limitations of the Model}
\label{sec:limitations}

The $Hab$ model and its reduced variants describe the average brightening behavior of the sample, not necessarily the behavior of any single comet. Individual lightcurves can depart from the population trend for identifiable physical reasons, causing that comet's behavior to deviate from the population-level model without necessarily invalidating it. The fitted parameters can also carry systematic biases from our specific dataset and reduction choices, which could shift the inferred $(a,b)$ distribution without appearing as outliers at all.

We selected comets that were discovered at $r_\mathrm{h} \gtrsim 5$~au, observed over a pre-perihelion range of $\Delta r_\mathrm{h} > 1$~au and that have $q < 6.6$~au. We additionally only considered observations from $r_\mathrm{h} > 2.5$~au. Comets discovered closer to the Sun, with larger perihelion distances, or observed over shorter arcs, are not represented here. Using the $1/a_0 < 4\times10^{-5}$~au$^{-1}$ threshold from Section~\ref{sec:intro}, 16 of the 23 comets are dynamically new; the sample is weighted toward DNCs, and while we found no clear difference in brightening behavior between DNCs and returning comets, that comparison relied on only seven returning comets, only four of which would still be classified as returning under the stricter $1/a_0 > 10^{-4}$~au$^{-1}$ threshold used by \citet{lacerda25}. 

All of our fits use pre-perihelion (inbound) data only; we do not test or validate any of the models presented here against post-perihelion lightcurves in this work. As noted in Section~\ref{sec:intro}, the post-perihelion brightening behavior of LPCs at large distances remains poorly characterized in the literature more broadly, and it is not yet known whether the same $(a,b)$ values, or even the same functional form, would describe a comet's outbound evolution. We are extending the LOOK sample to post-perihelion data and plan to address this question directly in future work.

Because our photometry uses a fixed $5\arcsec$ angular aperture, the volume of coma sampled grows with geocentric distance $\Delta$, biasing resolved comae to appear relatively brighter at large $\Delta$: flux scales closer to $\Delta^{-1}$ than the point-source $\Delta^{-2}$ of Equation~\ref{eqn:traditional}, though this can shift back toward $\Delta^{-2}$ if a tail extends beyond the aperture \citep{jewitt91}. We already substituted $\Delta^{-1}$ for $\Delta^{-2}$ scaling in all of our model fits (Equation~\ref{eqn:aper_correct}) to partially account for this; however, the transition between coma- and tail-dominated regimes is not captured by this single correction. Since this transition is not modeled explicitly, it is degenerate with the $r_\mathrm{h}$-dependence we fit, and would be absorbed into $a$ and $b$ rather than appearing as clear deviations in the residuals. Our reported $(a,b)$ values are therefore specific to this aperture choice.

Individual comets can also deviate from the population trend for reasons our models do not capture: outbursts, which are well documented in distant LPCs \citep[e.g.,][]{kelley22, hui19} and can bias a fit or produce a temporary offset from the trend at any stage; seasonal and projection effects, where a changing illuminated area or coma/tail geometry alters apparent brightness independent of activity \citep{weissman88, schleicher07a, kelley25}; and comets like C/2019 L3, whose rapid brightness surge beyond $r_\mathrm{h} \sim 3.7$~au departs strongly from the population trend (e.g., Fixed-$ab$ RMSE $= 0.54$~mag, the worst fit in the sample). These are reminders that our models describe the typical inbound evolution of an LPC discovered near 5--10~au, not a universal law.

An outburst or a surge will likely still be visible as an outlier relative to the population trend. For brightness predictions, where the goal is the likely range of activity rather than an exact forecast, the reduced models in Table~\ref{tab:recommendations} remain a practical tool available at discovery.

\section{Summary}
\label{sec:summary}

We compared the traditional constant activity-index model ($Hn$) to an empirical distance-dependent brightening model ($Hab$) and its anchored, reduced-parameter variants, using real pre-perihelion lightcurves for 23 long-period comets from the LCO Outbursting Objects Key (LOOK) project. We summarize our results below.

\vspace{10pt}
\textit{Characterizing distant brightening behavior:}
\begin{enumerate}
    \item For inbound photometry at $r_\mathrm{h} \gtrsim 2.5$~au, a constant activity index often cannot describe the full distant lightcurve; allowing the effective slope to vary with heliocentric distance substantially improves the fits.
    \item The empirical three-parameter $Hab$ model ($ar_\mathrm{h} + b$) provides a significantly better match to our distant lightcurves than the traditional $Hn$ model.
    \item The best-observed comets cluster in $(a,b)$ space ($A_1 = 0.40 \pm 0.09$, $B_1 = -1.78 \pm 0.68$), motivating one- and two-parameter variants (Fixed-$ab$, Fixed-$a$, Fixed-$b$, and Correlated-$ab$) anchored to these population statistics; the Bayesian information criterion favors the full three-parameter $Hab$ model over all four variants for the majority of objects, justifying the added complexity for characterization.
\end{enumerate}

\textit{Predicting future brightness:}
\begin{enumerate}
    \setcounter{enumi}{3}
    \item Short early segments poorly constrain $Hab$, so although it is best for full-curve characterization, it is comparatively poor for early-time magnitude forecasting.
    \item The constrained variants are more robust predictors: Fixed-$ab$ performs well even for very short initial segments, while the two-parameter variants can improve quickly with modest additional coverage ($\Delta r_\mathrm{h} \gtrsim 1$ -- 1.5~au). Correlated-$ab$ converges earliest among the two-parameter cases, with Fixed-$b$ close behind.
\end{enumerate}

\textit{Applicability and outlook:}
\begin{enumerate}
    \setcounter{enumi}{5}
    \item These models were fit only to pre-perihelion (inbound) data and should not be assumed to apply post-perihelion, where fading behavior remains poorly characterized. They also do not capture all pre-perihelion variability (e.g., outbursts, seasonal effects), and should not be relied on to describe activity inside the water-dominated regime ($r_\mathrm{h} \lesssim 2.5$~au), where none of the models tested here were fit or validated. All fits also assume that the brightness measured in our fixed 5$''$ angular aperture scales as $\Delta^{-1}$ (Equation~\ref{eqn:aper_correct}), so the reported $(a, b)$ values apply to photometry fit with the same scaling.
    \item Additional long-term lightcurves provided by LSST should tighten population constraints on $(a,b)$ and strengthen the priors behind reduced-parameter prediction for survey and mission applications.
\end{enumerate}

\section*{Acknowledgments}
The authors would like to sincerely thank the two anonymous reviewers, whose valuable feedback greatly strengthened this article. CEH was supported by an LSST-DA Catalyst Fellowship during this work, made possible through the support of Grant 62192 from the John Templeton Foundation to LSST-DA. CS acknowledges funding from the UK Space Agency. This work makes use of observations from the Las Cumbres Observatory global telescope network under proposal codes KEY2020B-009, LTP2023B-001, and LTP2025A-004. The raw frames and Basic Calibrated Data (BANZAI-processed) products are available at the LCO Science Archive at \url{https://archive.lco.global}. 

The views expressed in this article are those of the authors and do not reflect the official policy or position of the U.S.\ Naval Academy, Department of the Navy, the Department of Defense, or the U.S.\ Government.

During the preparation of this work the authors used a large language model for minor copy editing. After using this tool/service, the authors reviewed and edited the content as needed and take full responsibility for the content of the published article.

%

\vspace{5mm}
\facilities{LCOGT (1m Sinistro)}


\software{{\tt astropy} \citep{astropy13,astropy18,astropy22}, {\tt numpy} \citep{harris20}, {\tt pandas} \citep{mckinney10}, {\tt astroquery} \citep{astroquery19}, {\tt scipy} \citep{scipy20}, {\tt sbpy} \citep{mommert19}, JPL Horizons \citep{horizons}, {\tt matplotlib} \citep{matplotlib}, {\tt seaborn} \citep{waskom21}, NEOexchange \citep{neoexchange2023ascl}, BANZAI \citep{mccully18}, {\tt calviacat} \citep{kelley19}, SExtractor \citep{sextractor}
          }

\break
\appendix
\renewcommand{\thefigure}{A\arabic{figure}} 
\setcounter{figure}{0} 

\section{Model Ranking and Significance Visualizations}
\label{app:rankings}

\begin{figure*}[ht]
\centering
\includegraphics[width=\textwidth]{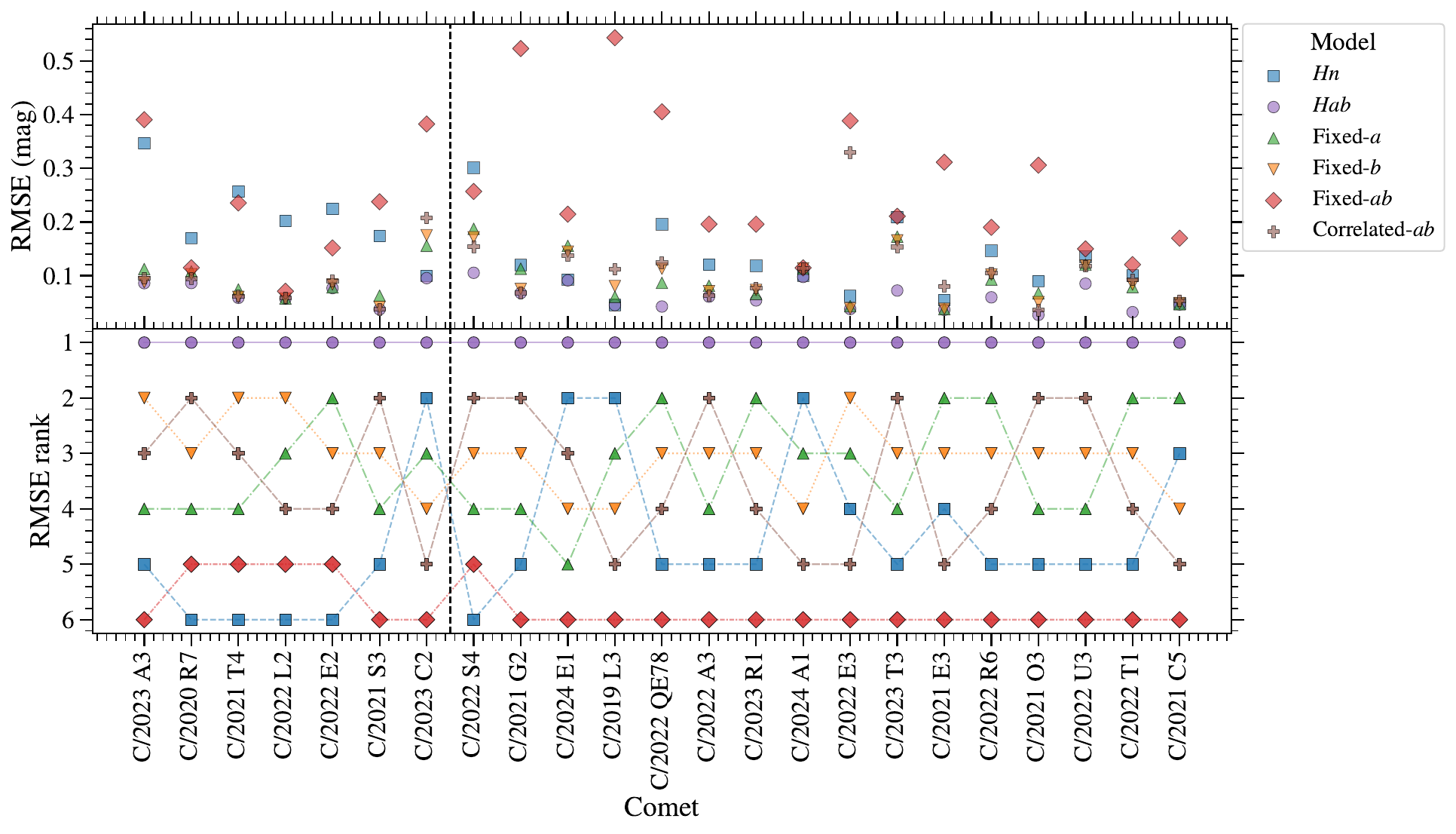}
\caption{RMSE (top) and RMSE rank (bottom) for each of the six models tested, for every comet in the sample. The seven comets with $\mathcal{R}>2$ are separated from the remaining sample by the vertical dashed line.}
\label{fig:rmse_rank}
\end{figure*}

\begin{figure*}[ht]
\centering
\includegraphics[width=\textwidth]{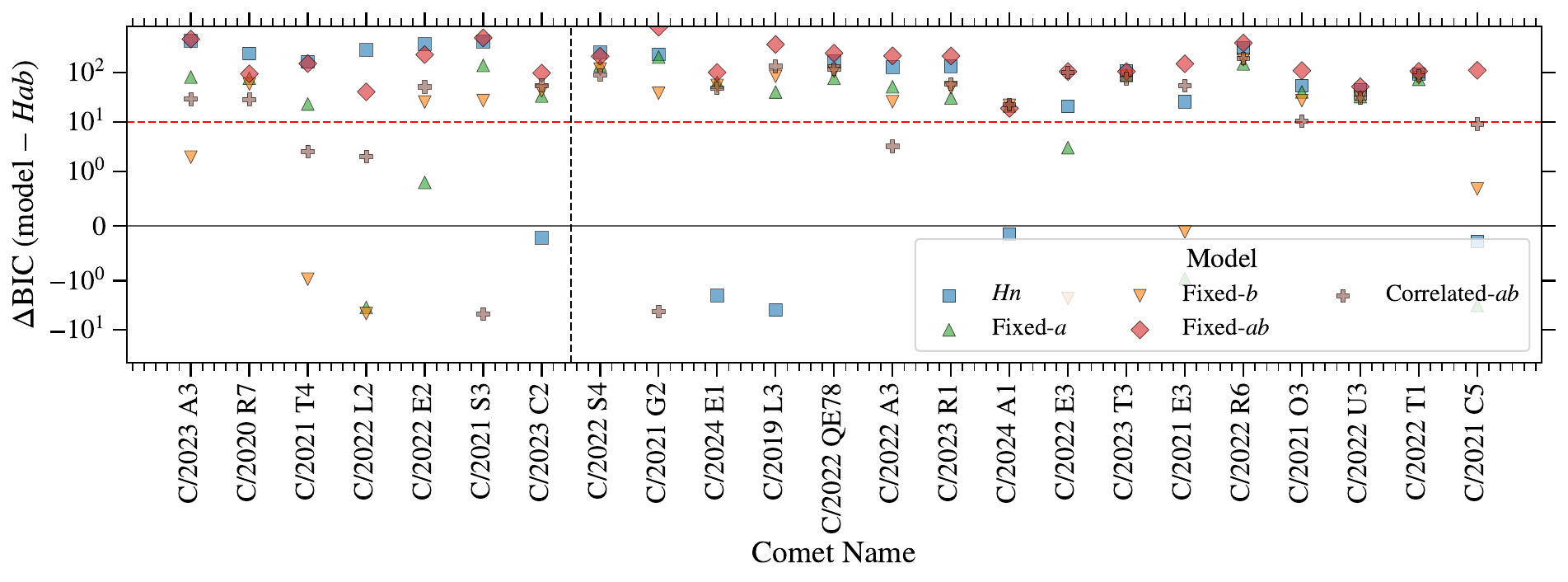}
\caption{$\Delta\mathrm{BIC} = \mathrm{BIC}_\mathrm{model} - \mathrm{BIC}_{Hab}$ for each of the five alternative models, for every comet in the sample, plotted on a symmetric-log ($y$-)scale so that negative and zero values are visible alongside the more extreme positive ones. The solid gray horizontal line marks $\Delta\mathrm{BIC}=0$, where negative values indicate that the alternative model is preferred over $Hab$; the dashed red horizontal line marks $\Delta\mathrm{BIC}=10$, the threshold for strong evidence in favor of $Hab$. The seven comets with $\mathcal{R}>2$ are separated from the remaining sample by the vertical dashed line.}
\label{fig:delta_bic}
\end{figure*}

Figure~\ref{fig:rmse_rank} shows the RMSE (top panel) and corresponding rank (bottom panel) of each model for every comet in the sample, as described in Section~\ref{sec:results}. Figure~\ref{fig:delta_bic} shows the $\Delta$BIC of each model relative to $Hab$ for every comet, with the $\Delta\mathrm{BIC} = 10$ threshold used to define strong evidence in Section~\ref{sec:results} marked for reference.

\FloatBarrier
\section{Predictive Test Diagnostics}
\label{app:predict_diagnostics}
\begin{figure*}[ht]
\centering
\includegraphics[width=\textwidth]{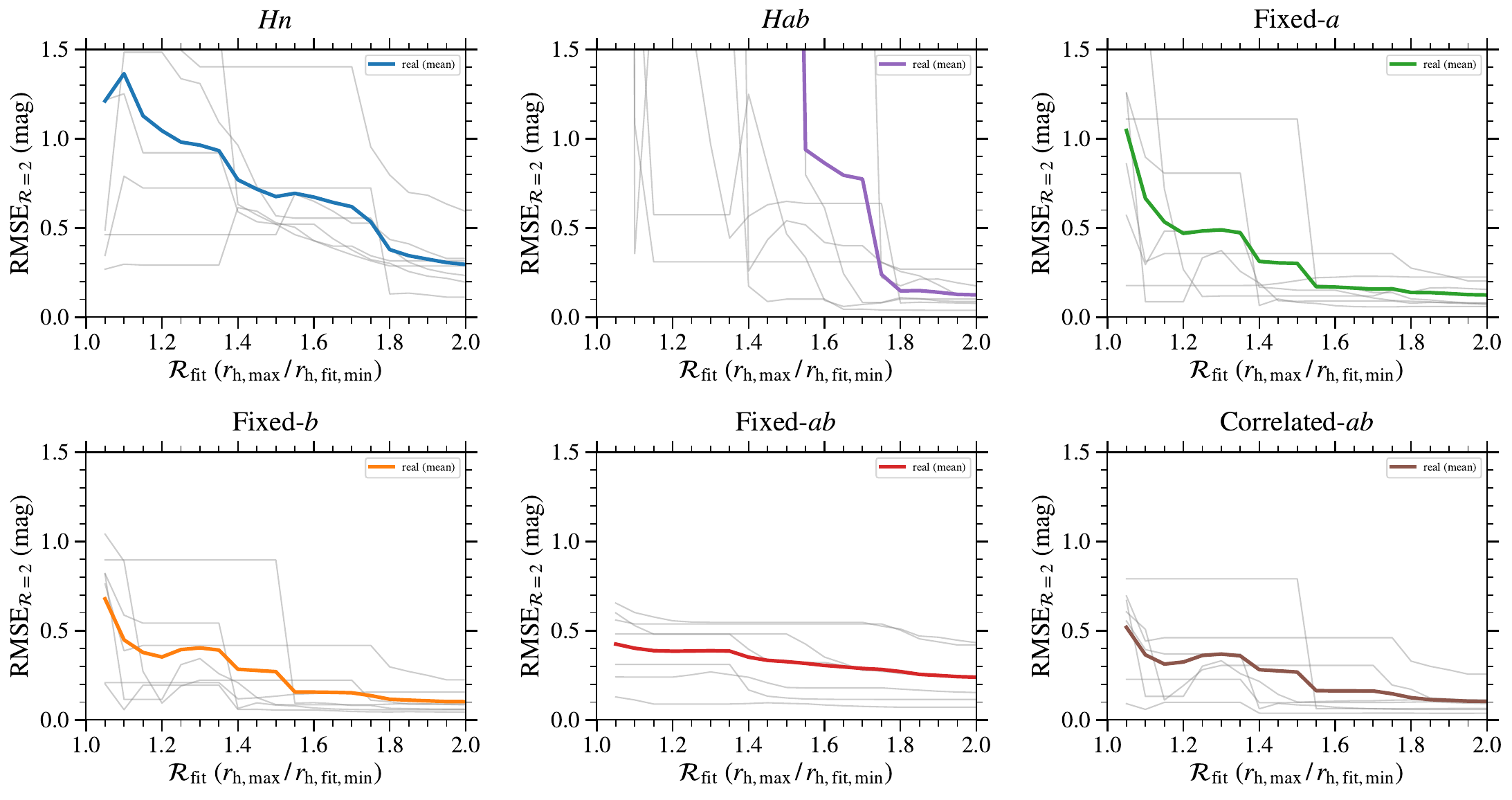}
\caption{Individual per-comet RMSE$_{\mathcal{R}=2}$ curves (gray) underlying the sample mean shown in Figure \ref{fig:rmse_predict}, one panel per model. Individual comets can show non-monotonic jumps where the fit-arc length does not grow smoothly with $\mathcal{R}_\mathrm{fit}$ (Figure \ref{fig:R_vs_deltarh}); these average out in the sample mean but are visible here directly, demonstrating that the modest structure in Figure \ref{fig:rmse_predict} reflects small-sample sampling behavior rather than instability in any single model's fit.}
\label{fig:rmse_permodel}
\end{figure*}

\begin{figure*}[ht]
\centering
\includegraphics[width=0.7\textwidth]{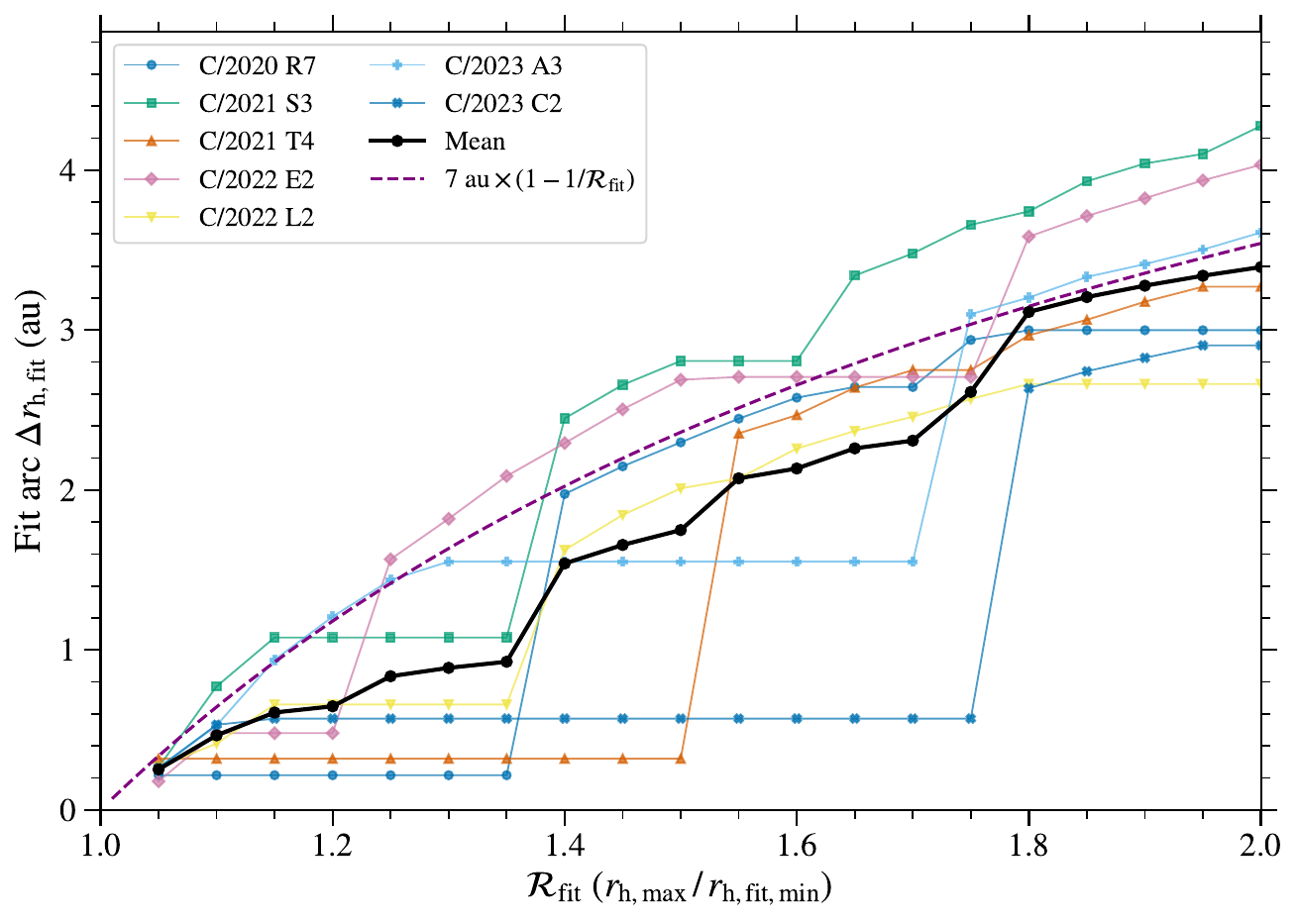}
\caption{Heliocentric distance range of the data actually used for each fit ($\Delta r_\mathrm{h,fit} = r_\mathrm{h,max} - r_\mathrm{h,fit,min}$) as a function of the ratio threshold $\mathcal{R}_\mathrm{fit}$, for each of the seven $\mathcal{R} > 2$ comets (colored lines) and the sample mean (black). The dashed purple line shows the smooth theoretical expectation, $\langle r_\mathrm{h,max}\rangle(1 - 1/\mathcal{R}_\mathrm{fit})$, evaluated at the sample-mean $r_\mathrm{h,max} \approx 7$ au; it is not a per-comet prediction, so comets with individual $r_\mathrm{h,max}$ above/below this average (\rev{5.97--8.58~au} across the sample) trace systematically above/below it. Within a given comet, the fit-arc length only increases when $\mathcal{R}_\mathrm{fit}$ crosses an observed epoch, producing the step-like behavior seen here rather than a smooth curve. This discreteness, not model instability, is the origin of the non-monotonic structure in Figures \ref{fig:rmse_predict} and \ref{fig:mag_predict}.}
\label{fig:R_vs_deltarh}
\end{figure*}

Figure \ref{fig:rmse_permodel} shows the individual per-comet RMSE$_{\mathcal{R}=2}$ curves underlying the sample-mean curves presented in Figure \ref{fig:rmse_predict} (Section \ref{sec:results_predict}), and Figure \ref{fig:R_vs_deltarh} shows the heliocentric distance range of data actually used for each fit as a function of $\mathcal{R}_\mathrm{fit}$ for each comet. Because only seven comets contribute to the sample mean, and each is sampled at its own irregular real cadence, the fit-arc length used at a given threshold grows in discrete steps as $\mathcal{R}_\mathrm{fit}$ crosses real observation epochs, rather than continuously (Figure \ref{fig:R_vs_deltarh}). This produces the modest non-monotonic structure visible in Figures \ref{fig:rmse_predict} and \ref{fig:mag_predict} around $\mathcal{R}_\mathrm{fit} \sim 1.2$ -- 1.4; Figure \ref{fig:rmse_permodel} confirms that this behavior arises from small-sample sampling granularity, not from instability in any individual model's predictions.

\FloatBarrier
\section{Model Fit Parameters}
\label{app:fit_params}
Table~\ref{tab:model_fits} lists the best-fit parameters for each comet and model described in Section~\ref{sec:hab_variants}.
\begin{longrotatetable}
\begin{deluxetable*}{l r@{$\,\pm\,$}l r@{$\,\pm\,$}l c r@{$\,\pm\,$}l r@{$\,\pm\,$}l r@{$\,\pm\,$}l c r@{$\,\pm\,$}l c r@{$\,\pm\,$}l r@{$\,\pm\,$}l c r@{$\,\pm\,$}l r@{$\,\pm\,$}l c r@{$\,\pm\,$}l r@{$\,\pm\,$}l r@{$\,\pm\,$}l}
\tablecaption{Model fits summary\label{tab:model_fits}}
\tablecolumns{32}
\tablewidth{0pt}
\tabletypesize{\scriptsize}
\setlength{\tabcolsep}{0.0175in}
\renewcommand{\arraystretch}{0.9}
\tablehead{
  \colhead{Comet} &
  \multicolumn{4}{c}{\textit{Hn}} &&
  \multicolumn{6}{c}{\textit{Hab}} &&
  \multicolumn{2}{c}{Fixed-$\mathit{ab}$} &&
  \multicolumn{4}{c}{Fixed-$\mathit{a}$} &&
  \multicolumn{4}{c}{Fixed-$\mathit{b}$} &&
  \multicolumn{6}{c}{\rev{Correlated-$\mathit{ab}$}} \\
  \cline{2-5} \cline{7-12} \cline{14-15} \cline{17-20} \cline{22-25} \cline{27-32}
  \colhead{Name} &
  \multicolumn{2}{c}{$H$} & \multicolumn{2}{c}{$n$} &&
  \multicolumn{2}{c}{$H$} & \multicolumn{2}{c}{$a$} & \multicolumn{2}{c}{$b$} &&
  \multicolumn{2}{c}{$H$} &&
  \multicolumn{2}{c}{$H$} & \multicolumn{2}{c}{$b$} &&
  \multicolumn{2}{c}{$H$} & \multicolumn{2}{c}{$a$} &&
  \multicolumn{2}{c}{$H$} & \multicolumn{2}{c}{$a$} & \multicolumn{2}{c}{$b$} \\
}
\startdata
C/2019 L3 & 2.67 & 0.05 & 5.84 & 0.03 &  & 2.30 & 0.60 & -0.04 & 0.06 & 6.24 & 0.63 &  & 11.61 & 0.06 &  & 6.84 & 0.06 & 1.40 & 0.04 &  & 9.85 & 0.03 & 0.69 & 0.01 &  & 13.89 & 0.06 & 1.07 & 0.02 & -6.04 & 0.10 \\
C/2020 R7 & 10.92 & 0.08 & 2.27 & 0.05 &  & 13.53 & 0.12 & 0.28 & 0.01 & -0.65 & 0.13 &  & 14.43 & 0.01 &  & 14.69 & 0.05 & -1.94 & 0.03 &  & 14.57 & 0.02 & 0.38 & 0.01 &  & 14.26 & 0.02 & 0.34 & 0.01 & -1.40 & 0.04 \\
C/2021 C5 & 8.01 & 0.17 & 5.15 & 0.10 &  & 11.22 & 1.74 & 0.28 & 0.15 & 1.91 & 1.75 &  & 16.35 & 0.03 &  & 12.58 & 0.16 & 0.54 & 0.10 &  & 14.88 & 0.07 & 0.60 & 0.01 &  & 17.70 & 0.07 & 0.85 & 0.02 & -4.61 & 0.14 \\
C/2021 E3 & 8.87 & 0.10 & 4.10 & 0.07 &  & 12.51 & 0.57 & 0.53 & 0.08 & -0.47 & 0.71 &  & 15.04 & 0.05 &  & 11.61 & 0.07 & 0.66 & 0.05 &  & 13.55 & 0.03 & 0.69 & 0.01 &  & 18.96 & 0.18 & 1.45 & 0.05 & -8.45 & 0.30 \\
C/2021 G2 & 6.63 & 0.08 & 3.70 & 0.04 &  & 12.06 & 0.26 & 0.21 & 0.01 & -0.41 & 0.20 &  & 12.38 & 0.04 &  & 17.00 & 0.07 & -4.16 & 0.04 &  & 13.87 & 0.02 & 0.28 & 0.01 &  & 12.33 & 0.01 & 0.22 & 0.01 & -0.62 & 0.01 \\
C/2021 O3 & 11.15 & 0.23 & 4.03 & 0.16 &  & 20.07 & 0.62 & 1.43 & 0.10 & -7.61 & 0.80 &  & 17.22 & 0.06 &  & 13.66 & 0.17 & 0.75 & 0.12 &  & 15.59 & 0.06 & 0.72 & 0.01 &  & 22.52 & 0.13 & 1.81 & 0.04 & -10.75 & 0.23 \\
C/2021 S3 & 7.31 & 0.11 & 3.81 & 0.06 &  & 13.26 & 0.11 & 0.31 & 0.01 & -1.21 & 0.10 &  & 13.32 & 0.02 &  & 15.03 & 0.04 & -2.70 & 0.02 &  & 13.94 & 0.01 & 0.34 & 0.01 &  & 13.21 & 0.01 & 0.31 & 0.01 & -1.17 & 0.01 \\
C/2021 T4 & 9.30 & 0.18 & 3.85 & 0.12 &  & 15.03 & 0.19 & 0.49 & 0.02 & -2.11 & 0.19 &  & 15.15 & 0.03 &  & 13.98 & 0.05 & -1.02 & 0.03 &  & 14.71 & 0.02 & 0.46 & 0.01 &  & 15.51 & 0.02 & 0.53 & 0.01 & -2.59 & 0.03 \\
C/2022 A3 & 8.00 & 0.18 & 4.69 & 0.11 &  & 18.12 & 0.61 & 0.83 & 0.05 & -5.32 & 0.60 &  & 15.45 & 0.02 &  & 12.89 & 0.12 & -0.15 & 0.08 &  & 14.51 & 0.04 & 0.54 & 0.01 &  & 16.39 & 0.03 & 0.69 & 0.01 & -3.64 & 0.07 \\
C/2022 E2 & 7.33 & 0.14 & 3.53 & 0.08 &  & 14.40 & 0.20 & 0.43 & 0.01 & -2.78 & 0.18 &  & 12.89 & 0.01 &  & 13.93 & 0.05 & -2.35 & 0.03 &  & 13.26 & 0.02 & 0.36 & 0.01 &  & 12.78 & 0.01 & 0.34 & 0.01 & -1.37 & 0.02 \\
C/2022 E3 & 8.75 & 0.11 & 3.49 & 0.09 &  & 11.69 & 0.50 & 0.69 & 0.12 & -1.09 & 0.77 &  & 13.59 & 0.08 &  & 10.47 & 0.08 & 0.82 & 0.06 &  & 12.13 & 0.03 & 0.79 & 0.01 &  & 23.43 & 3.36 & 2.90 & 0.85 & -17.73 & 5.45 \\
C/2022 L2 & 11.63 & 0.11 & 2.08 & 0.07 &  & 14.95 & 0.10 & 0.39 & 0.01 & -1.81 & 0.11 &  & 14.83 & 0.01 &  & 15.06 & 0.03 & -1.93 & 0.02 &  & 14.92 & 0.01 & 0.39 & 0.01 &  & 14.71 & 0.02 & 0.36 & 0.01 & -1.54 & 0.03 \\
C/2022 QE78 & 4.63 & 0.32 & 4.71 & 0.15 &  & 24.88 & 0.62 & 0.66 & 0.02 & -9.66 & 0.44 &  & 12.07 & 0.05 &  & 16.89 & 0.14 & -3.98 & 0.07 &  & 13.74 & 0.07 & 0.30 & 0.01 &  & 12.46 & 0.02 & 0.26 & 0.01 & -0.88 & 0.04 \\
C/2022 R6 & 1.07 & 0.28 & 6.40 & 0.13 &  & 33.93 & 1.12 & 0.86 & 0.03 & -15.06 & 0.73 &  & 12.23 & 0.01 &  & 16.46 & 0.18 & -3.65 & 0.08 &  & 13.54 & 0.06 & 0.33 & 0.01 &  & 12.59 & 0.02 & 0.31 & 0.01 & -1.16 & 0.03 \\
C/2022 S4 & 10.12 & 0.18 & 3.78 & 0.11 &  & 20.64 & 0.36 & 0.89 & 0.03 & -6.96 & 0.36 &  & 15.99 & 0.02 &  & 14.86 & 0.11 & -1.06 & 0.07 &  & 15.53 & 0.04 & 0.47 & 0.01 &  & 16.40 & 0.03 & 0.55 & 0.01 & -2.70 & 0.06 \\
C/2022 T1 & 14.51 & 0.23 & 1.15 & 0.15 &  & 27.96 & 0.72 & 1.67 & 0.09 & -14.56 & 0.84 &  & 16.42 & 0.02 &  & 17.75 & 0.18 & -2.64 & 0.12 &  & 16.99 & 0.09 & 0.31 & 0.01 &  & 15.62 & 0.15 & 0.16 & 0.04 & -0.26 & 0.28 \\
C/2022 U3 & 6.50 & 0.30 & 5.35 & 0.16 &  & 46.80 & 4.62 & 2.12 & 0.24 & -28.29 & 3.86 &  & 15.51 & 0.02 &  & 14.13 & 0.26 & -1.02 & 0.14 &  & 15.03 & 0.09 & 0.45 & 0.01 &  & 15.66 & 0.03 & 0.49 & 0.01 & -2.31 & 0.09 \\
C/2023 A3 & 6.41 & 0.12 & 4.01 & 0.07 &  & 12.08 & 0.12 & 0.51 & 0.01 & -2.11 & 0.13 &  & 12.50 & 0.03 &  & 10.86 & 0.04 & -0.79 & 0.02 &  & 11.76 & 0.02 & 0.48 & 0.01 &  & 12.80 & 0.01 & 0.57 & 0.01 & -2.86 & 0.02 \\
C/2023 C2 & 9.64 & 0.08 & 3.41 & 0.05 &  & 8.54 & 0.61 & -0.13 & 0.07 & 4.70 & 0.71 &  & 14.73 & 0.06 &  & 13.14 & 0.12 & -0.71 & 0.08 &  & 14.06 & 0.06 & 0.50 & 0.01 &  & 15.46 & 0.09 & 0.65 & 0.03 & -3.36 & 0.17 \\
C/2023 R1 & 6.11 & 0.20 & 4.82 & 0.10 &  & 23.10 & 0.94 & 0.63 & 0.03 & -7.67 & 0.69 &  & 14.11 & 0.02 &  & 17.02 & 0.11 & -3.20 & 0.05 &  & 15.07 & 0.04 & 0.33 & 0.01 &  & 14.17 & 0.01 & 0.30 & 0.01 & -1.12 & 0.03 \\
C/2023 T3 & 10.73 & 0.30 & 3.14 & 0.19 &  & 29.42 & 0.98 & 2.01 & 0.10 & -17.50 & 1.08 &  & 15.71 & 0.03 &  & 14.47 & 0.25 & -0.99 & 0.16 &  & 15.14 & 0.10 & 0.48 & 0.01 &  & 16.45 & 0.11 & 0.64 & 0.04 & -3.28 & 0.23 \\
C/2024 A1 & 10.11 & 0.12 & 3.18 & 0.07 &  & 7.05 & 1.49 & -0.25 & 0.12 & 6.22 & 1.47 &  & 15.18 & 0.01 &  & 14.98 & 0.14 & -1.66 & 0.08 &  & 15.11 & 0.06 & 0.41 & 0.01 &  & 15.21 & 0.03 & 0.42 & 0.01 & -1.86 & 0.09 \\
C/2024 E1 & 8.97 & 0.12 & 3.87 & 0.06 &  & 7.78 & 0.83 & -0.06 & 0.04 & 4.86 & 0.69 &  & 15.09 & 0.03 &  & 16.61 & 0.20 & -2.55 & 0.10 &  & 15.73 & 0.08 & 0.35 & 0.01 &  & 15.07 & 0.02 & 0.31 & 0.01 & -1.21 & 0.06 \\
\enddata
\begin{tablenotes}[para,flushleft]
\item The free parameter values determined for each model for each comet using a least-squares fit. All uncertainties are $1\sigma$. Goodness-of-fit comparisons (RMSE and $\Delta$BIC) for these fits are presented in Section~\ref{sec:results_bestfit} and Table~\ref{tab:model_fit_stats}.
\end{tablenotes}
\end{deluxetable*}
\end{longrotatetable}

\onecolumngrid

\clearpage




\end{CJK*}
\end{document}